\documentclass[fleqn,usenatbib]{mnras}

\usepackage{newtxtext,newtxmath}

\usepackage[T1]{fontenc}

\DeclareRobustCommand{\VAN}[3]{#2}
\let\VANthebibliography\thebibliography
\def\thebibliography{\DeclareRobustCommand{\VAN}[3]{##3}\VANthebibliography}

\usepackage{graphicx}	
\usepackage{multirow}
\usepackage{listings}
\usepackage{xcolor}

\newcommand{\teff}{${T}_{\rm eff}$}
\newcommand{\logg}{$\mathrm{log(g)}$}
\newcommand{\plax}{$\pi$}

\title[System parameters of three short period CVs]{System parameters of three short period cataclysmic variable stars}

\author[J. Wild, S. Littlefair]{
J. F. Wild$^1$,
S. P. Littlefair$^1$,
R. P. Ashley$^{2,5}$,
E. Breedt$^3$,
A. J. Brown$^1$,
V. S. Dhillon$^{1,6}$,
\newauthor
M. J. Dyer$^1$,
M. J. Green$^{2,4}$,
P. Kerry$^1$,
T. R. Marsh$^2$,
S. G. Parsons$^1$,
D. I. Sahman$^1$,
\\
$^1$ Department of Physics and Astronomy, University of Sheffield, Sheffield, S3~7RH, United Kingdom
\\
$^2$ Astronomy and Astrophysics Group, Department of Physics, University of Warwick, Coventry, CV4 7AL, United Kingdom
\\
$^3$ Institute of Astronomy, University of Cambridge, Madingley Road, Cambridge CB3 0HA, UK
\\
$^4$ School of Physics and Astronomy, Faculty of Exact Sciences, Tel Aviv University, Tel Aviv 69978, Israel \\
$^5$ Isaac Newton Group of Telescopes, Apartado de Correos 321, Santa Cruz de La Palma, E-38700, Spain
$^6$ Instituto de Astrofisica de Canarias, E38205 La Laguna, Tenerife, Spain
}

\date{Accepted XXX. Received YYY; in original form ZZZ}

\pubyear{2020}

\begin{document}
\label{firstpage}
\pagerange{\pageref{firstpage}--\pageref{lastpage}}
\maketitle
\begin{abstract}
Using photometric ULTRACAM observations of three new short period cataclysmic variables, we model the primary eclipse lightcurves to extract the orbital separation, masses, and radii of their component stars. 
We find donor masses of $0.060\pm0.008 {\rm M_\odot}$, $0.042\pm0.001 {\rm M_\odot}$, and $0.042\pm0.004 {\rm M_\odot}$, two being very low-mass sub-stellar donors, and one within $2\sigma$ of the hydrogen burning limit. 
All three of the new systems lie close to the empirical evolutionary sequence that has emerged from observations of the last decade.
We briefly re-evaluate the long-standing discrepancy between observed donor mass and radius data, and theoretical CV evolutionary tracks.
By looking at the difference in the observed period at each mass and the period predicted by the modelled evolutionary sequences, we qualitatively examine the form of excess angular momentum loss that is missing from the models below the period gap.
We show indications that the excess angular momentum loss missing from CV models grows in importance relative to gravitational losses as the period decreases. 
Detailed CV evolutionary models are necessary to draw more quantitative conclusions in the future.
\end{abstract}




\section{Introduction}
\label{sect:introduction}

Cataclysmic Variable (CV) stars are binary systems, containing white dwarf primary stars, and low mass 
companion stars \citep{hellier2001}.
Generally, the white dwarf is the more massive of the two, but CVs with the majority of the system mass contained in the companion star are possible \citep{ritter2003}.
The two stars orbit close enough that the companion completely fills its Roche lobe and the outer layers of its envelope are gradually stripped from its surface, falling towards the white dwarf around which an accretion disc forms. The companion has its mass transferred to the primary, so is referred to as the donor star (e.g. \citet{warner1995}).

CVs evolve from long to short orbital periods, driven by the contraction of the donor in response to mass loss \citep{patterson1984}. For CVs with orbital periods less than 6 hours, mass loss is a consequence of angular momentum loss (AML) from the binary.
AML in CVs is generally considered to result from two mechanisms: gravitational wave braking \citep{paczynski1981}, and magnetic braking \citep{rappaport1983}. CVs with long periods, more than $\sim$3 hours, exhibit both magnetic braking and gravitational wave braking, but below this period magnetic braking appears to cease and the donor contracts to its equilibrium radius \citep{spruit1983, Davis2008}. This causes the donor to detach from the Roche lobe and mass transfer stops, leading to a period gap where CVs are not observed \citep{Kolb1998, hellier2001, knigge2006}. The stars move closer together through gravitational losses, until at $\sim$2.2 hours the donor reconnects with its Roche lobe \citep{Davis2008} and mass transfer resumes as a short-period CV, though with a significantly reduced transfer rate \citep{paczynski1981, Rappaport1982, kolb99}. The CV eventually evolves through a period minimum when the thermal timescale of the donor becomes comparable to its mass loss timescale.
Once beyond the period minimum, the donor expands in response to mass loss, allowing it to sustain mass transfer as it retreats and leading to a widening of the orbit \citep{paczynski1981, Rappaport1982, knigge2006}.

The observed location of the period minimum has been difficult to reproduce in evolutionary models (see \citealt{zorotovic2020} for a review of this history), and the most common explanation of this discrepancy is an extra source of AML over the traditional gravitational wave and magnetic losses \citep{King1995, knigge11}.
The donor mass and radius are expected to be a valuable diagnostic for CV evolution, as they should trace the long-term average mass-loss of the system \citep{knigge11}. Observations have so far produced an evolutionary sequence with little scatter between donor mass and radius, or between donor mass and orbital period, implying that CVs quickly converge on a singular evolutionary path \citep{McAllister2019}.

A physically motivated solution for missing AML was proposed by \citet{King1995}, in which angular momentum is lost as a consequence of mass transfer, and hence is called consequential AML, or CAML. \citet{Schreiber2016} suggest this is caused by mass ejection from nova outbursts, making AML a function of the white dwarf mass and accretion rate. With some tuning, this idea is able to solve three long-standing problems in CV evolution: low observed CV space density (e.g. \citealt{britt2015}), the missing observations of systems below the period gap \citep{Kolb1993a, knigge2006}, and the observed high CV white dwarf masses (e.g. \citealt{McAllister2019}). However, there is not yet any direct evidence for this theory.

While promising, CAML is not the only potential extra source of AML. 
The CV field has long made the assumption that magnetic braking either fully, or almost fully, ceases at the period gap \citep{mcdermot1989, taam1989}, leaving AML dominated by gravitational wave braking. However, it is unlikely that magnetic braking fully stops, and the strength of the remainder is unknown, only assumed to be negligible.
Magnetic braking requires a strong magnetic field to couple to a significant stellar wind, but \citet{Garraffo2018} recently suggested that the structure of the magnetic field has a strong influence on its ability to drive a stellar wind, based on work by \citet{taam1989}. A more complex field will produce fewer open field lines, which are required to eject particles from the system and carry away angular momentum. 
\citet{morin2010} find a wide range of field complexities in M dwarf stars, which is difficult to reconcile with the single, unified track driven by magnetic braking found by \citet{knigge11}.
However, as solitary low mass stars with the high rotation rates of CV donors are extremely rare, the \citet{morin2010} data do not cover the relevant region of the parameter space. It is feasible that the rapid rotation rates of CV donor stars stabilise the magnetic fields enough on thermal timescales to produce the observed singular tracks. 
At least some residual magnetic braking is likely to be present below the period gap, but the question of how significant it is to the AML history of the CV remains.

The best probe for the AML history of a CV is the donor mass and radius evolution over orbital period \citep{knigge11}. However, direct measurements of masses and radii of the components of the very low mass ratio CVs found at short periods are hard won and few in number, and \citet{McAllister2019} report only a handful of such systems.
By modelling the eclipse of the white dwarf - a technique established by \citet{wood1986} and further developed by \citet{Savoury2011} and \citet{McAllister2017} - we characterise three new CVs. Our method is described in \S\ref{sect:modelling the cv} 

We characterise three recently identified CVs: ASASSN-16kr, ASASSN-17jf, and CRTS SSS11126 J052210-350530. 
These systems have been chosen for their short periods, and prior observations of each system are summarised below. Table~\ref{table:system locations and ephemerides} and \S\ref{sect:prior observations} summarise their observational information.

\begin{table*}
    \centering
    \label{table:system locations and ephemerides}
    \caption{Summary of objects observed for this work. Given magnitudes are the approximate magnitudes out of eclipse observed in this work. $T_0$ and $P$ are the ephemerides calculated in \S\ref{sect:ephemeris data}. Parallax (\plax) is as measured by Gaia DR2 \citep{lindegren2018, Luri2018, Gaia2016, Gaia2018}. $N_\mathrm{ecl}$ is the number of ULTRACAM eclipse observations used in this analysis.}
    \begin{tabular}{ccc|cc|cccc|c|c}
        \hline
        System & RA & Dec & $T_0$, & $P_\mathrm{orb}$, & \multicolumn{4}{c}{Approx. System Magnitude} & \plax,  & $N_\mathrm{ecl}$\  \\
         & & & BMJD, TDB (err) & days (err) & $u'$\ & $g'$\ & $r'$\ & $i'$\ & mas & \\
        \hline\hline
        ASASSN-16kr & 22:05:59.48 & -34:14:33.9    & $58635.424328(3)$ & $0.061285932(1)$ & $19.1$ & $19.5$ & $19.8$ & $20.1$ & $6.230\pm0.266$ & 7 \\
        ASASSN-17jf & 20:29:17.13 & -43:40:19.8    & $58756.50523(1)$  & $0.0567904(7)$   & $20.7$ & $20.1$ & $20.3$ & $-$    & $3.494\pm1.127$ & 3 \\
        SSSJ0522-3505 & 05:22:09.67 & -35:05:30.3 & $58799.52170(1)$  & $0.06219343(1)$  & $19.1$ & $19.0$ & $19.3$ & $-$    & $1.214\pm0.323$ & 3 \\
        \hline
    \end{tabular}
\end{table*}

\subsection{Prior observations}
\label{sect:prior observations}

\subsubsection{ASASSN-16kr}
\label{sect:ASASSN-16kr prior observations}
ASASSN-16kr, a.k.a. MASTER J220559.40-341434.9, was discovered by the All-Sky Automated Survey for Supernovae (ASASSN) on 11 September 2016, and observed by the MASTER network on the 19th (ATel \#9509 and \#9510), both at $\sim 14^{\rm th}$ magnitude.
Initially classified as an SS Cyg type object due to its low outburst amplitude (vsnet-alert \#20189), subsequent observations confirmed eclipses and superhumping behaviour (vsnet alerts \#20190, \#20196, \#20206; \citealt{kato2017}). 

Time-resolved photometery detected superhumps and eclipses, and \citet{kato2017} calculated an orbital period of $0.0612858\pm0.0000003$ days, and a superhump period of $0.061999\pm0.000067$ days. \citet{Kato2009a} demonstrated that superhump periods vary systematically, and can be categorised into stages: stage A, an initial growth stage with a long period; stage B, a developed stage with a varying period; and stage C, with a shorter and more constant period. This system is noted by \citet{kato2017} as being in the transition from stage B to stage C, though this is noted as possibly being due to a suspect measurement a the start of the outburst they observed.

\subsubsection{ASASSN-17jf}
\label{sect:ASASSN-17jf prior observations}
ASASSN-17jf was confirmed as eclipsing by Berto Monard (vsnet \#21257) between 14 and 17 July 2017. The system was initially observed with a mean unfiltered magnitude of $\sim15.5$ outside eclipse, with an eclipse depth of $\sim1$\ magnitude. From these observations, an orbital period of $0.0578\pm 0.0003$\ days, and a rough superhump period of $0.0565$\ days was derived.

\subsubsection{CRTS SSSJ0522-3505 J052210-350530}
\label{sect:SSSJ0522-3505 prior observations}
CRTS SSSJ0522-3505 J052210-350530, hereafter SSSJ0522-3505, was first observed by the CRTS on 28 February 2005, and as recently as 11 November 2019 \citep{drake2009}. These data show high variability, and outbursts $\sim$6 months apart. High time resolution lightcurves taken by \citet{paterson2019} show an eclipse depth of $\sim1.5$\ magnitudes and an orbital period of $0.0622 \pm 0.0005$\ days.

\section{Observations and data reduction}
\label{sect:observations}

Observations were taken with ULTRACAM \citep{dhillon2007}, mounted on the 3.58m New Technology Telescope (NTT) in La Silla, Chile. ULTRACAM is a three-colour camera capable of observing these $\sim20$th magnitude systems at a time resolution of a few seconds, with a signal/noise ratio high enough to resolve the various components of the eclipse. 

Observations were taken on several nights in four observing runs spanning from 13 October 2018 to 29 January 2020. Table~\ref{table:observations} summarises these observations. A full discussion of calibrating the data is given in Appendix \ref{sect:photometric extraction and calibration}. Briefly, instrument signature removal and aperture photometery was performed using the HiPERCAM pipeline software\footnote{\url{http://www.vikdhillon.staff.shef.ac.uk/hipercam/resources.html}}, and flux calibration used nearby comparison stars in conjunction with known flux secondary standards.

\begin{table*}
    \centering
    \caption{Journal of Observations. Each eclipse is imaged in three colours simultaneously by ULTRACAM mounted on the NTT. SDSS-like filters are denoted by subscript {\it reg}, and upgraded, higher throughput filters are denoted by subscript {\it sup}; see \S\ref{sect:colour correction method} for details.}
    \label{table:observations}
    \begin{tabular}{c | ccc | p{2cm}c |ccc}
        \hline
        System & Date & Start time & Stop time & $T_\mathrm{ecl}$\ & Ecl. $\mathrm{N^o}$\ & Filters & Flux standard & Airmass \\
         & & UTC & UTC & BMJD,~TDB,~(err)  & & & used & \\
        \hline \hline
        ASASSN-16kr 
        & 2018-10-13$^\dagger$ & 02:34:58 & 03:15:43 & 58404.131217(3) & -3774 & $u_{\rm reg}, g_{\rm reg},r_{\rm reg}$ & G 27-45 & 1.04-1.10 \\
        & 2018-10-16$^\dagger$ & 04:25:49 & 04:59:32 & 58407.1955(2)   & -3724 & $u_{\rm reg}, g_{\rm reg},r_{\rm reg}$ & G 27-45 & 1.33-1.50\\
        & 2018-10-17$^\dagger$ & 02:24:23 & 04:26:57 & 58408.114806(4), 58408.176(1) & -3709, -3708 & $u_{\rm reg}, g_{\rm reg},r_{\rm reg}$ & G 27-45 & 1.05-1.35 \\
        & 2019-09-27 & 23:56:59 & 00:27:17 & 58754.012610(3) & 1935 & $u_{\rm sup}, g_{\rm sup},r_{\rm sup}$ & SA 114 548 & 1.11-1.17\\
        & 2019-09-29 & 00:48:44 & 01:37:34 & 58755.054468(3) & 1952 & $u_{\rm sup}, g_{\rm sup},r_{\rm sup}$ & SA 114 548 & 1.02-1.06\\
        & 2019-09-30 & 03:21:59 & 04:02:34 & 58756.157613(4) & 1970 & $u_{\rm sup}, g_{\rm sup},r_{\rm sup}$ & SA 114 548 & 1.03-1.09\\
        \hline
        ASASSN-17jf
        & 2019-09-28 & 01:41:39 & 03:04:00 & 58754.12003(2) & -42 & $u_{\rm sup}, g_{\rm sup},r_{\rm sup}$& SA 114 548 & 1.05-1.16\\
        & 2019-09-30 & 02:16:18 & 02:46:29 & 58756.10769(1) & -7  & $u_{\rm sup}, g_{\rm sup},r_{\rm sup}$& SA 114 548 & 1.10-1.14\\
        & 2019-10-01 & 04:08:56 & 04:38:37 & 58757.18671(1) & 12  & $u_{\rm sup}, g_{\rm sup},r_{\rm sup}$& SA 114 548 & 1.40-1.55\\
        \hline
        SSSJ0522-3505 
        & 2019-09-29 & 08:12:53 & 09:00:37 & 58755.364361(6) & -710 & $u_{\rm sup}, g_{\rm sup},r_{\rm sup}$ & SA 114 548 & 1.01-1.05 \\
        & 2019-10-01 & 08:01:32 & 08:42:20 & 58757.35456(1)  & -678 & $u_{\rm sup}, g_{\rm sup},r_{\rm sup}$ & SA 114 548 & 1.02-1.06 \\
        & 2020-01-29 & 04:07:50 & 05:02:36 & 58877.20128(5)  & 1249 & $u_{\rm sup}, g_{\rm sup},i_{\rm sup}$ & BD -210910 & 1.19-1.39 \\
        \hline
    \end{tabular}
    {$^\dagger$Calibration of these data use the uncorrected standard magnitudes provided in \citet{smith2002}, without the colour corrections described in \S\ref{sect:colour correction method} and \S\ref{sect:flux calibrating the lightcurve}.}
\end{table*}

\section{Modelling the CV}
\label{sect:modelling the cv}

To determine the system parameters for the three CVs in this study, the eclipse lightcurves were modelled. This method is more frequently applicable in CVs than the more traditional approach of using spectroscopic eclipsing binaries, since the donor star is rarely directly visible. Compared to using the superhump period excess to estimate the mass ratio \citep{patterson2005, knigge2006}, lightcurve modelling requires few assumptions. However, it does require precise alignment of the system and so is not possible for a large fraction of CVs.

Several excellent discussions of the technique exist in the literature (e.g. \citealt{wood1986, Savoury2011, McAllister2017, McAllister2019}), though we summarise key elements of the approach here. Four assumptions are made: the bright spot originates where a ballistic trajectory from the donor meets the outer edge of the accretion disc, the white dwarf obeys a theoretical mass-radius relationship, the white dwarf is unobscured by the accretion disc or other sources of intra-system material, and the donor exactly fills its Roche lobe. Most of these assumptions are considered robust, though the visibility of the white dwarf been called into question by \citet{Spark2015}. Since the white dwarf radius is inferred from the duration of ingress/egress, complicating structures like a surface layer of accreted material could lead to an inaccurate white dwarf radius, and hence mass. However, system parameters from lightcurve modelling agree with other methods \citep{tulloch2009, copperwheat2012, savoury2012}, suggesting that this is not normally an issue.
The model for one eclipse is described by 18 parameters:
\begin{enumerate}
    \item white dwarf, disc, bright spot, and donor fluxes, $F_\mathrm{(WD,\ disc,\ BS,\ donor)}$;
    \item mass ratio, $q = \frac{M_\mathrm{donor}}{M_\mathrm{WD}}$;
    \item white dwarf eclipse width, in units of phase, $\Delta\phi$;
    \item scaled white dwarf radius, $R_\mathrm{WD}/a$;
    \item white dwarf limb darkening coefficient, $u_\mathrm{ld}$;
    \item scaled outer disc radius, $R_\mathrm{disc}/a$;
    \item disc surface profile exponent;
    \item seven parameters describing the bright spot behaviour;
    \item an eclipse phase offset, $\phi_0$;
\end{enumerate}
where $a$\ is orbital separation between the white dwarf and donor star. The seven bright spot parameters describe its brightness profile and beaming, location on the rim of the accretion disc, and emission angle, but are not physically motivated. For details, see \citet{Savoury2011}.

In addition, there are three nuisance parameters, that set the timescale and amplitude of a Gaussian process that describes flickering. These parameters are common to all eclipses for a system.

\subsection{Lightcurve fitting procedure}
\label{sect: lightcurve fitting}

We extend the lightcurve fitting model used by \citet{McAllister2019}, adopting a hierarchical approach to slightly reduce model complexity. 

\begin{figure}
    \centering
    \includegraphics[width=.85\columnwidth ]{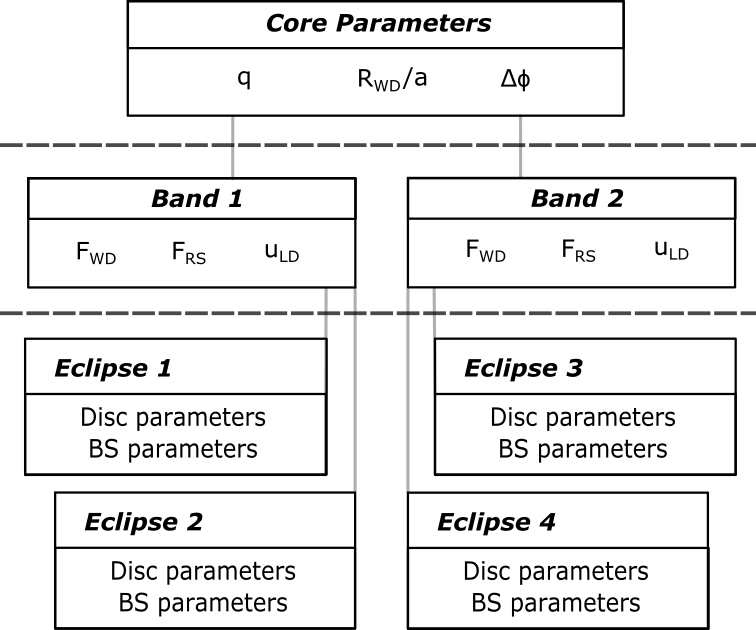}
    \caption{The hierarchical structure of the lightcurve model. Parameters are inherited downwards, to produce an eclipse at the `leaves' of the tree, e.g. Eclipse 3 inherits the parameters of Band 2, which in turn inherits the Core parameters. $\mathrm{F_{WD, RS}}$\ represent the fluxes of the white dwarf and donor star, and $\mathrm{U_{LD}}$\ is the limb darkening coefficient of the white dwarf.}
    \label{fig:hierarchical_model}
\end{figure}

Changes in the disc radius and brightness profile, and bright spot parameters can mean that the same CV has a significantly different eclipse lightcurve at different times, making it difficult to justify averaging together many eclipses, as features can become smeared out and uninformative. In the worst-case scenario, all 18 parameters would be independently variable for each eclipse, in each band. However, by sharing some parameters between eclipses and bands, this large number of free parameters is slightly reduced, and the posterior of some parameters can be informed by multiple eclipses. \citet{McAllister2017} share $q,\ R_{\rm WD}/a$, and $\Delta\phi$ between eclipses, and we extend that concept by organising the model into a hierarchical tree structure, a schematic of which is shown in Figure~\ref{fig:hierarchical_model}. 

The top level of the model provides the core parameters, which are unchanging between all observing bands and constant across our observations: $q,\ R_\mathrm{WD}/a$, and $\Delta\phi$. We assume the white dwarf and donor fluxes do not change on the timescale of our observations, and so these variables, along with the limb darkening coefficient of the white dwarf, are shared between all eclipses observed with the same filters. The bottom level holds parameters that can vary quickly enough to change between eclipses, i.e. parameters describing the accretion disc and bright spot. By handling parameters this way, we maximise the amount of data informing important variables, for example, white dwarf fluxes and $q$. We also somewhat reduce the number of free parameters, which aids slightly in model fitting, but the chief justification for the hierarchical approach is that it ensures consistency between eclipses - something not guaranteed when fitting eclipses individually.

As more eclipses are added, the number of dimensions in parameter space that must be explored increases. For illustration, the model for ASASSN-17jf has 3 eclipses across 3 bands, plus 3 Gaussian process parameters, resulting in 87 free parameters that must be optimised simultaneously. To find the most likely set of lightcurve parameters in this very large space, an ensemble MCMC fitting code was used. The MCMC uses the \texttt{emcee} implementation of an ensemble sampler and parallel tempering \citep{foreman2012} to aid convergence to a global minimum despite the large parameter space, as described in \citet{McAllister2019}.

\subsection{Conversion to physical parameters}
\label{sect:physical_params conversion}

By capturing eclipses in multiple filters, preferably simultaneously, we can extract white dwarf colours from the eclipse fitting. Model white dwarf cooling tracks from \citet{Bergeron1995} list the absolute magnitudes of white dwarfs of a given \teff\ and \logg, and we fit these to the observed white dwarf fluxes, along with two nuisance parameters: parallax, \plax; and interstellar extinction, E(B-V). 
For E(B-V), the IRSA extinction maps were used to inform the prior, providing a maximum allowed value; uniform priors between zero and the maximum E(B-V)  were used.
A Gaussian prior on \plax\ based on Gaia data was used \citep{lindegren2018, Luri2018, Gaia2016, Gaia2018}. The priors used for \logg\ and \teff\ were more complicated, and are outlined in \S\ref{sect:method WD atmosphere fits}.

To calculate SI values for system parameters, we employ the technique developed by \citet{wood1986}.
White dwarfs follow well-understood cooling tracks that relate the stars' \teff, $R_{\rm WD}$, and $M_{\rm WD}$. We have an estimate for \teff\ as described above, so for an initial guess of the white dwarf mass, the cooling track provides a corresponding white dwarf radius. The relations we use are taken from \citet{wood1995} and \citet{panei2000}, which each cover a different range of $M_{\rm WD}$.

Eclipse modelling gives us a mass ratio, so the $M_{\rm WD}$\ guess can be used to calculate the total mass of the system, $M_{\rm T}$. $M_{\rm T}$\ and $P$, via Keplers' third law, gives the orbital separation. Using the $R_{\rm WD}/a$\ from eclipse modelling, $R_{\rm WD}$\ can be calculated. If the original guess for  $M_{\rm WD}$ is correct, the resulting $R_{\rm WD}$ will be consistent with the value obtained from the cooling track, allowing the correct white dwarf mass to be found.

Once the white dwarf mass has been found, we can calculate $a$ and re-scale the parameters produced by eclipse modelling to SI units. The following list of important system parameters is produced:
\begin{enumerate}
    \item white dwarf mass and radius, $M_\mathrm{WD}, R_\mathrm{WD}$
    \item the \logg\ corresponding to (i)
    \item white dwarf \teff
    \item donor mass and radius, $M_\mathrm{donor}, R_\mathrm{donor}$
    \item white dwarf - donor separation, $a$
    \item orbital velocities, $k_\mathrm{WD}, k_\mathrm{donor}$
\end{enumerate}
Note that \textit{two} values of \logg\ are produced in this process, one from fitting the white dwarf fluxes to model atmospheres, and one from combining the \teff\ estimate with lightcurve parameters.

\section{Results}
\label{sect:results general}

For all three systems eclipse modelling gave good results, each lightcurve being well-modelled with small residuals -- for a catalogue of the fits, see Appendix~\ref{appendix:lightcurves}, and Figure~\ref{fig:ASASSN-16kr example lightcurves} for an example. 
The Gaussian processes describing flickering in the systems were consistent with little to no variability, as almost all the scatter in the flux residuals could be fully described by the uncertainty in flux measurement. 

\begin{figure*}
    \centering
    \includegraphics[width=\textwidth, trim={0.5 28.5cm 0.5 0}, clip]{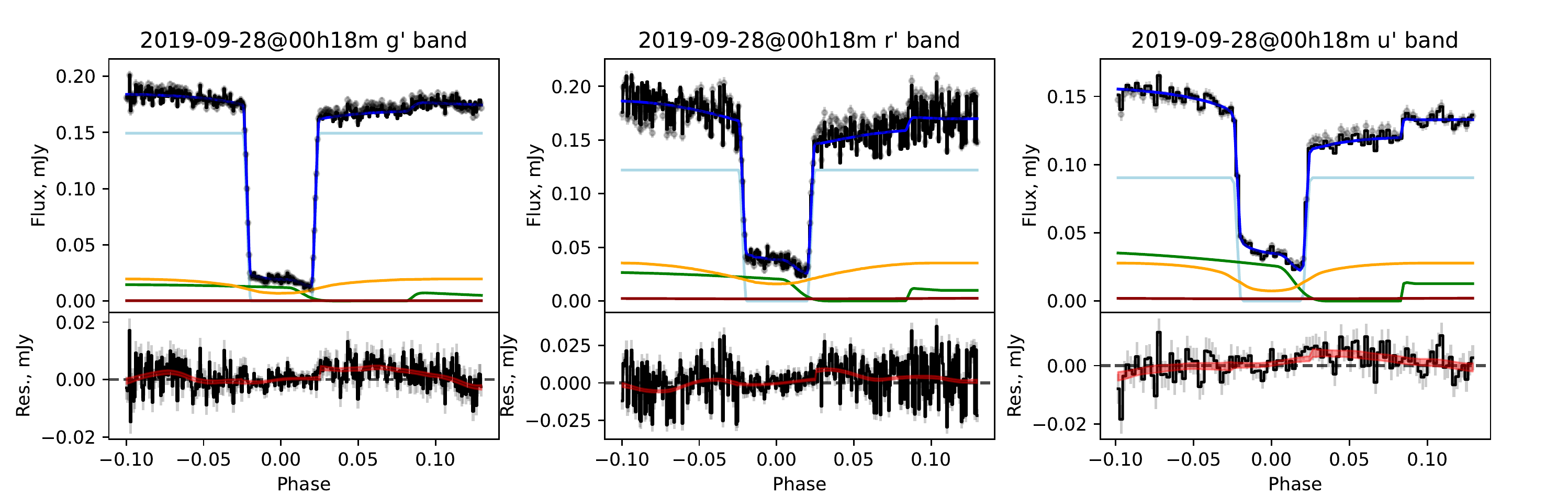}
    \caption{ASASSN-16kr example lightcurve models. {\it Top}:~grey points are the observed flux; black line is the observed flux, with the mean Gaussian process sample subtracted; the dark blue line is the mean lightcurve model, and the blue band is the standard deviation on this in the MCMC chain. The components of the model are also shown: the light blue line is the white dwarf flux, green line is the bright spot, orange line is the disc, and the red line is the donor. {\it Bottom}:~The residuals between the data and model are plotted as the black line, with grey error bars. The Gaussian process 1-sigma region is shown as a red band. A catalogue of all such fits in this work is given in Appendix~\ref{appendix:lightcurves}.}
    \label{fig:ASASSN-16kr example lightcurves}
\end{figure*}

\begin{table*}
    \centering
    \caption{The system parameters found for each system in this work.}
    \label{table:system_parameters}
    \begin{tabular}{c|c|c|c}
        \hline \\
        \textbf{System Name:}      & \textbf{ASASSN-16kr}    & \textbf{ASASSN-17jf}  & \textbf{SSSJ0522-3505} \\
        \hline \hline \\
        $M_\mathrm{WD}/M_\odot$    & $0.952\pm0.018$         & $0.669\pm0.031$        & $0.760\pm0.023$ \\
        $R_\mathrm{WD}/R_\odot$    & $0.0083\pm0.0002$       & $0.0120\pm0.0004$      & $0.0112\pm0.0003$ \\
        $M_\mathrm{donor}/M_\odot$ & $0.042\pm0.001$         & $0.060\pm0.008$        & $0.042\pm0.004$ \\
        $R_\mathrm{donor}/R_\odot$ & $0.105\pm0.002$         & $0.112\pm0.004$        & $0.105\pm0.004$ \\
        $q$                        & $0.044\pm0.002$         & $0.085\pm0.006$        & $0.055\pm0.003$ \\
        \hline
        $a/R_\odot$,               & $0.653\pm0.005$         & $0.567\pm0.009$        & $0.614\pm0.007$  \\
        $i$\                       & $86.4\pm0.4$            & $83.7\pm0.5$           & $83.8\pm0.3$  \\
        $K_\mathrm{WD}$, km/s      & $22.7\pm1.5$            & $39.5\pm4.2$           & $26.0\pm1.8$  \\
        $K_\mathrm{donor}$, km/s   & $515\pm3$               & $462\pm5$              & $470\pm4$  \\
        \hline
        \teff, kK                   & $10-12$           & $8-13$          & $\sim25$  \\
        \logg$, {\rm cgs}$        & $8.55\pm0.03$           & $8.15\pm0.05$          & $8.22\pm0.04$  \\
        \hline
    \end{tabular}
\end{table*}

\subsection{White dwarf atmosphere fits}
\label{sect:method WD atmosphere fits}

The two values of \logg\ produced by modelling -- the first from fitting the white dwarf fluxes to model atmospheres, and the second from combining \teff\ and $P$\ with the lightcurve parameters -- did not fall within $1\sigma$\ of each other in any of our systems. 
In ASASSN-17jf and SSSJ0522-3505, the white dwarf atmosphere fit converged close to the minimum surface gravity allowed by the coverage of our models, \logg$=7.0$. 
The second \logg, from lightcurve fitting, indicated values for each system of $8.10\pm0.04$ and $8.30\pm0.03$, respectively.
When analysing ASASSN-16kr, flux fitting gave a more reasonable \logg$=8.21\pm0.13$, but the second \logg\ still gave a significantly higher \logg$=8.59\pm0.03$, a difference of $\sim3\sigma$.

\begin{figure}
    \centering
    \includegraphics[width=\columnwidth, trim={0 0mm 0 0},clip]{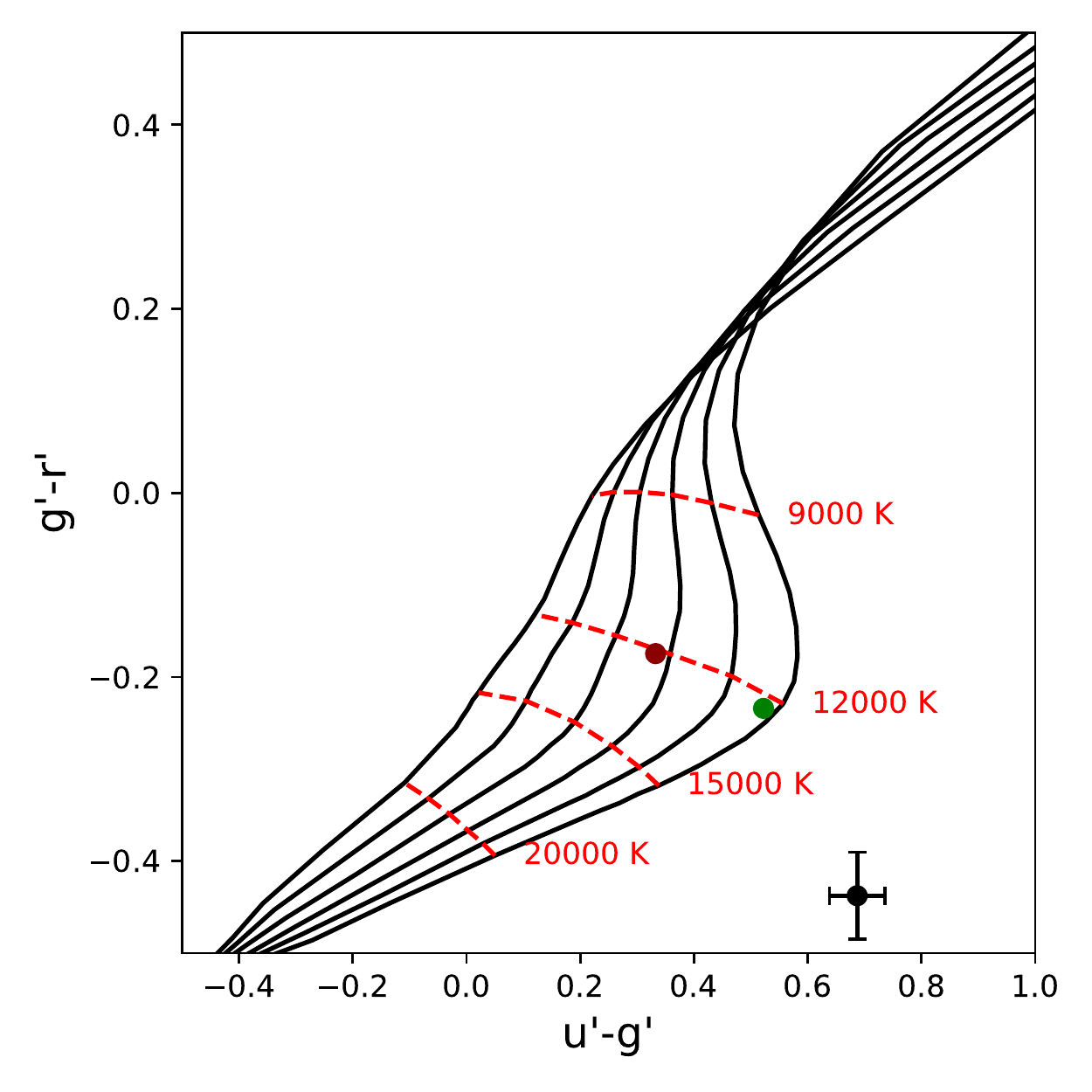}
    \caption{The white dwarf model atmosphere fits for ASASSN-17jf. Green circle: Best fit with uniform prior on \logg. Red circle: Best fit with the prior \logg$=8.10\pm0.04$. The observations are shown as the black point and error bars. Solid black lines are white dwarf model cooling tracks, increasing in \logg\ to the left. Red dashed lines are isothermal tracks for different \logg.}
    \label{fig:ASASSN-17jf colours}
\end{figure}

\begin{figure}
    \centering
    \includegraphics[width=\columnwidth, trim={0 0mm 0 0},clip]{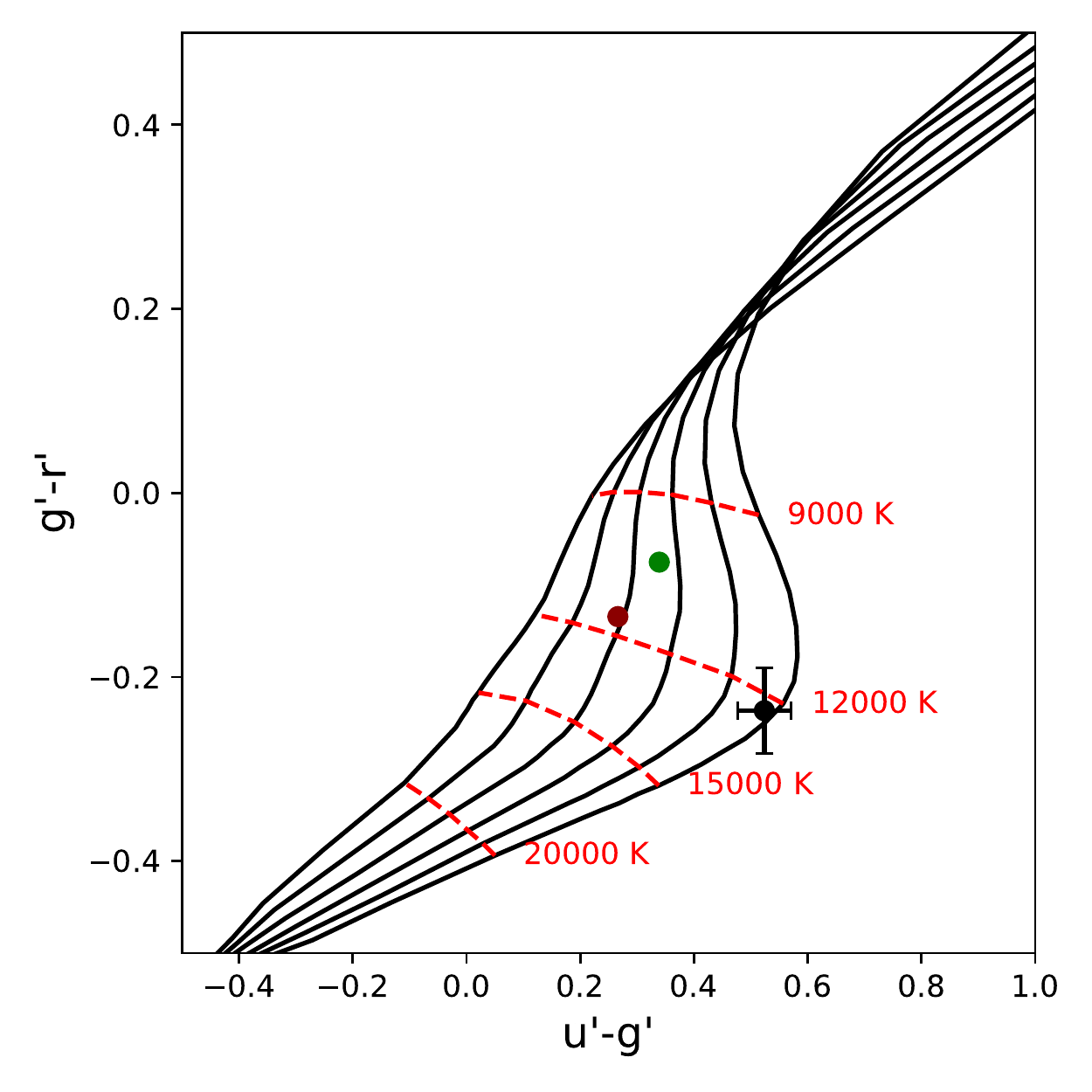}
    \caption{The white dwarf model atmosphere fits for ASASSN-16kr. The red circle is the best fit with a prior of \logg$=8.52\pm0.02$. Symbols are the same as Figure~\ref{fig:ASASSN-17jf colours}.}
    \label{fig:ASASSN-16kr colours}
\end{figure}

\begin{figure}
    \centering
    \includegraphics[width=\columnwidth, trim={0 0 0 0}, clip]{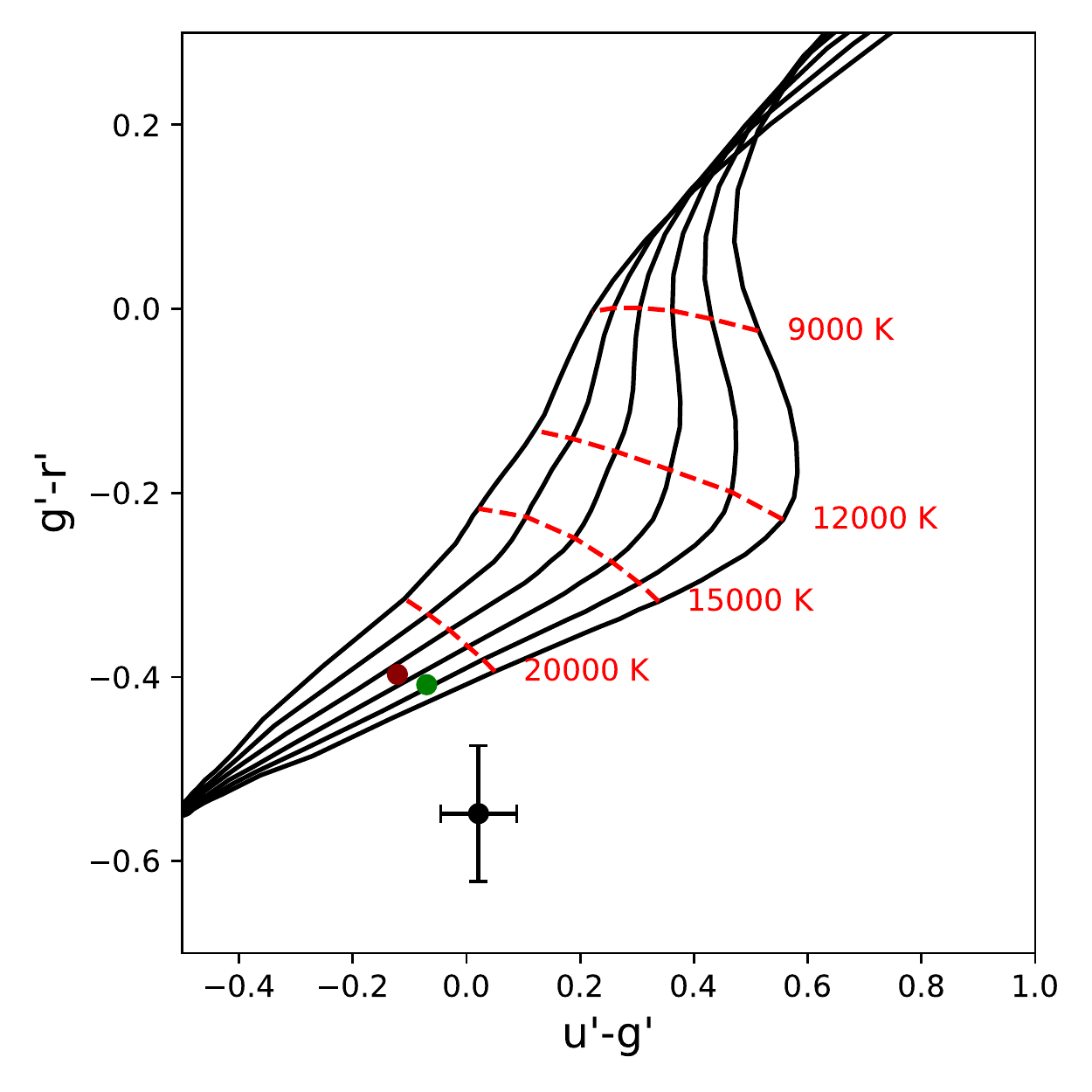}
    \caption{The white dwarf model atmosphere fits for SSSJ0522-3505. The red circle is the best fit with a prior of \logg$=8.28\pm0.04$. Symbols are the same as Figure~\ref{fig:ASASSN-17jf colours}.}
    \label{fig:SSSJ0522-3505 colours}
\end{figure}

This is concerning, as the two \logg\ should be consistent with one another for each system.
Comparison of our measured white dwarf colours to the \citet{Bergeron1995} model grids in Figures \ref{fig:ASASSN-17jf colours}, \ref{fig:ASASSN-16kr colours}, and \ref{fig:SSSJ0522-3505 colours}, reveals that the measured colours of the white dwarfs lie outside the colour space of the models. This is the origin of the discrepancies in \logg\ obtained with the two methods for ASASSN-17jf and SSSJ0522-3505, but ASASSN-16kr appears consistent with the leftmost cooling track. However, the observed flux of a white dwarf of this radius is too high for the observed Gaia parallax, pushing the model fits to smaller, higher gravity model atmospheres.

A possible cause for this issue would be an error in photometric calibration, causing a corresponding error in white dwarf fluxes. We do not believe this to be a problem, for the reasons explained in \S\ref{sect:flux calibrating the lightcurve}.
Inspection of the figures in Appendix~\ref{appendix:lightcurves} also rules out poor lightcurve fits as the cause of this problem. The most plausible explanation for the fact that our measured white dwarf fluxes do not lie inside the model grids, is that the change in brightness during white dwarf ingress/egress is contaminated by an additional source of light, for example a boundary layer close to the white dwarf surface. The implications of this for our system parameters is discussed in \S\ref{sect:impure white dwarf discussion}.

\begin{figure}
    \centering
    \includegraphics[width=\columnwidth, trim={0cm 6.5cm 0cm 0cm}, clip]{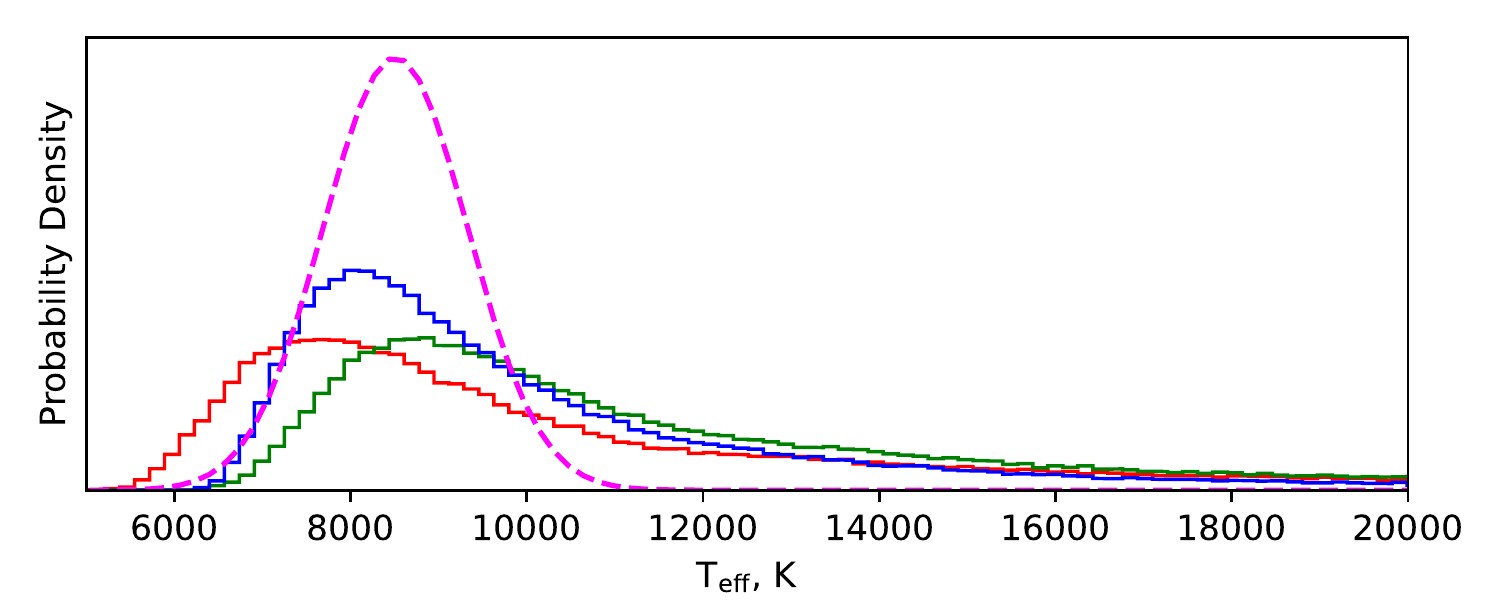}
    \includegraphics[width=\columnwidth, trim={0cm 6.5cm 0cm 0cm}, clip]{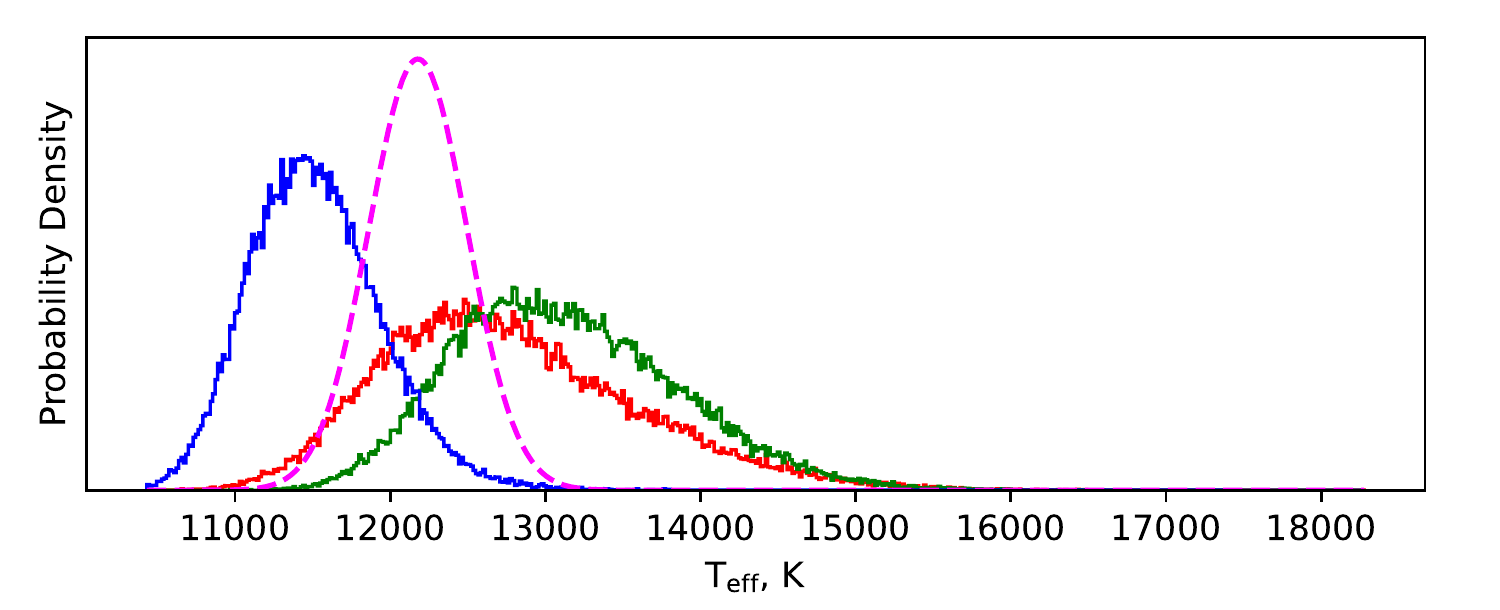}
    \includegraphics[width=\columnwidth, trim={0cm 6.5cm 0cm 0cm}, clip]{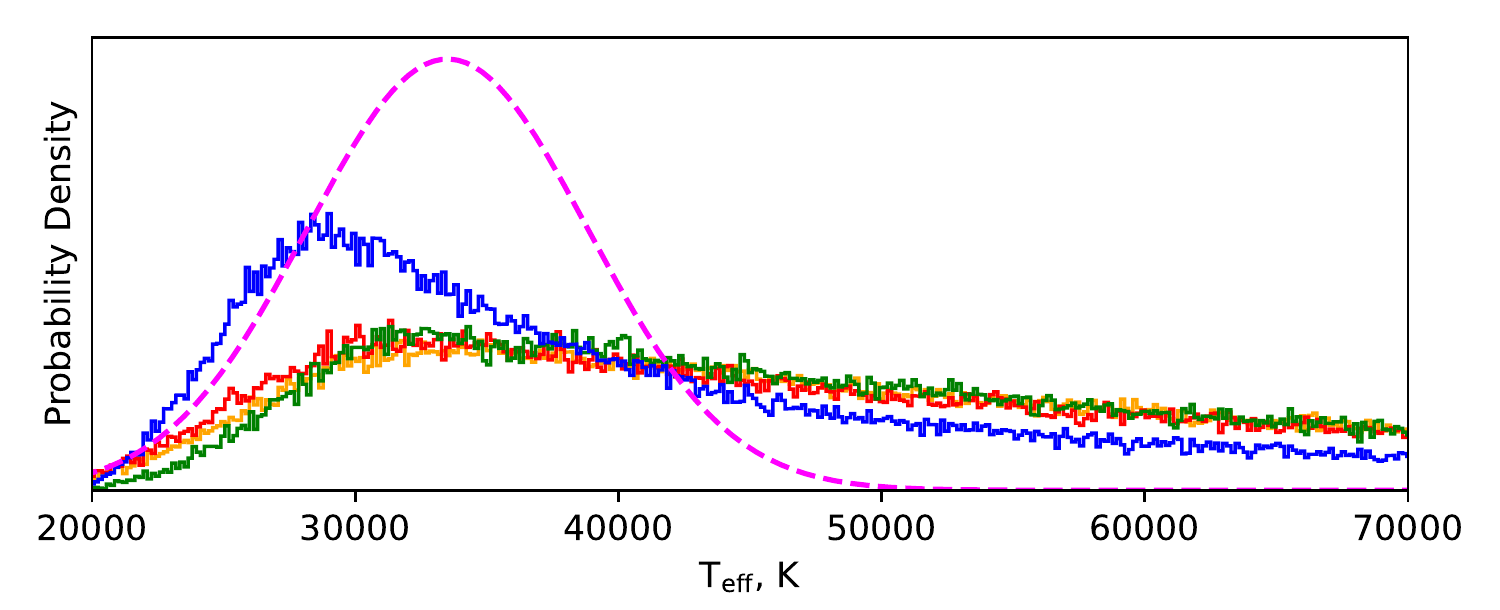}
    \caption{The result of fitting white dwarf model atmospheres to each photometric band independently. Blue solid line: $u'$ band, Green solid line: $g'$ band, Red solid line: $r'$ band. The joint distribution between all bands is characterised in each case by the best fit Gaussian (magenta dashed lines). \textit{Top}: ASASSN-17jf, joint \teff$=8330\pm780\rm\ K$; \textit{Middle}: ASASSN-16kr, joint \teff$=12150\pm300\rm\ K$; \textit{Bottom}: SSSJ0522-3505, joint \teff$=33300\pm5200 \rm\ K$. }
    \label{fig:gamma fits}
\end{figure}

That our white dwarf colours do not lie on the model grids also raises questions about the accuracy of our white dwarf temperatures. To try and quantify the impact on \teff\, we performed two additional fits to the white dwarf fluxes. In one approach we fit fluxes in all bands, but used a Gaussian prior on \logg\ using the estimate from the lightcurve modelling. In a second approach we fit the white dwarf flux in each band independently using the same prior on \logg\ and the Gaia prior on \plax. Since these independent fits use no colour information, E(B-V) is only constrained by the prior, but we retain it as a nuisance parameter and marginalise our \teff\ estimate over E(B-V).  Figure~\ref{fig:gamma fits} shows the \teff\ posteriors from the individual fits for the three systems.

From Figure \ref{fig:gamma fits}, we can see that there is little sign of a consistent discrepancy over the three observed CVs. The $u'$ band in ASASSN-16kr and SSSJ0522-3505 suggests a cooler temperature than the other bands, but lies in between the $r'$ and $g'$ in ASASSN-17jf.


\subsubsection{White dwarf temperature fits}
\label{sect:white dwarf temperature report}

Each approach gives a different distribution for \teff.
To avoid confusion, we do not report the results of each individual fit, but summarise the overall temperature ranges for each system. 

ASASSN-16kr \teff\ estimates ranged from 10200K to 12150K, and ASASSN-17jf estimates from 8330K to  12710K. 
The SSSJ0522-3505 fits that used all four observed fluxes both converged on $\sim22700$K, but the single-flux fits all resulted in wide posterior distributions covering $25000 - 90000$K, with very weak peaks in the $\sim30000 - 50000$K range, seen in Figure~\ref{fig:gamma fits}.

In all three systems, the figures we report in Table~\ref{table:system_parameters} are the \teff\ produced by the constrained \logg\ fit with all fluxes simultaneously. 
The \logg\ reported are the values found from the lightcurve parameters.

\subsection{System Parameters}
\label{sect:system parameters}

We note that the effect of the uncertain white dwarf temperatures on the system parameters, such as $M_{\rm wd}$, is negligible. For example, increasing \teff\ for ASASSN-17jf from 8000K to 12000K only changes $M_{\rm WD}$ by $0.001M_\odot$, compared to our statistical uncertainty of $0.031 M_\odot$. Even a large uncertainty in \teff\ only has a minor impact on the system parameters; for example a change in the WD temp for SSSJ0522-3505 from $10000$K to $20000$K only changes $M_{\rm WD}$ by $0.02 M_\odot$, comparable with the measurement uncertainty. The system parameters are reported in Table~\ref{table:system_parameters}.

ASASSN-16kr has a recorded superhump period, and now also a robust $q$ measurement. It can therefore be used to calibrate the superhump period excess, $\epsilon$ vs. $q$ relationship, as done in \citet{McAllister2019}, though with a more extreme mass ratio system than was available to them. The system was not confidently classed as exhibiting stage B or C stage superhumps, so we look at the results for both stages. Assuming the CV was in stage B, we calculate $q_B = 0.059\pm0.007$; assuming stage C and using the relevant relation from \citet{McAllister2019}, we calculate $q_C = 0.068\pm0.012$. In both cases, the estimated $q_\mathrm{B,C}$ is $\sim 2 \sigma$ higher than the observed value of $q = 0.044\pm0.002$. While a $2 \sigma$ difference is not a highly significant discrepancy, this may be preliminary evidence that the $\epsilon - q$ relation may over estimate $q$ for CVs at short periods, which has been suspected for some time \citep{pearson2007, knigge11}.


\section{Discussion}
\label{sect:discussion}

\begin{figure*}
    \centering
    \includegraphics[width=\textwidth]{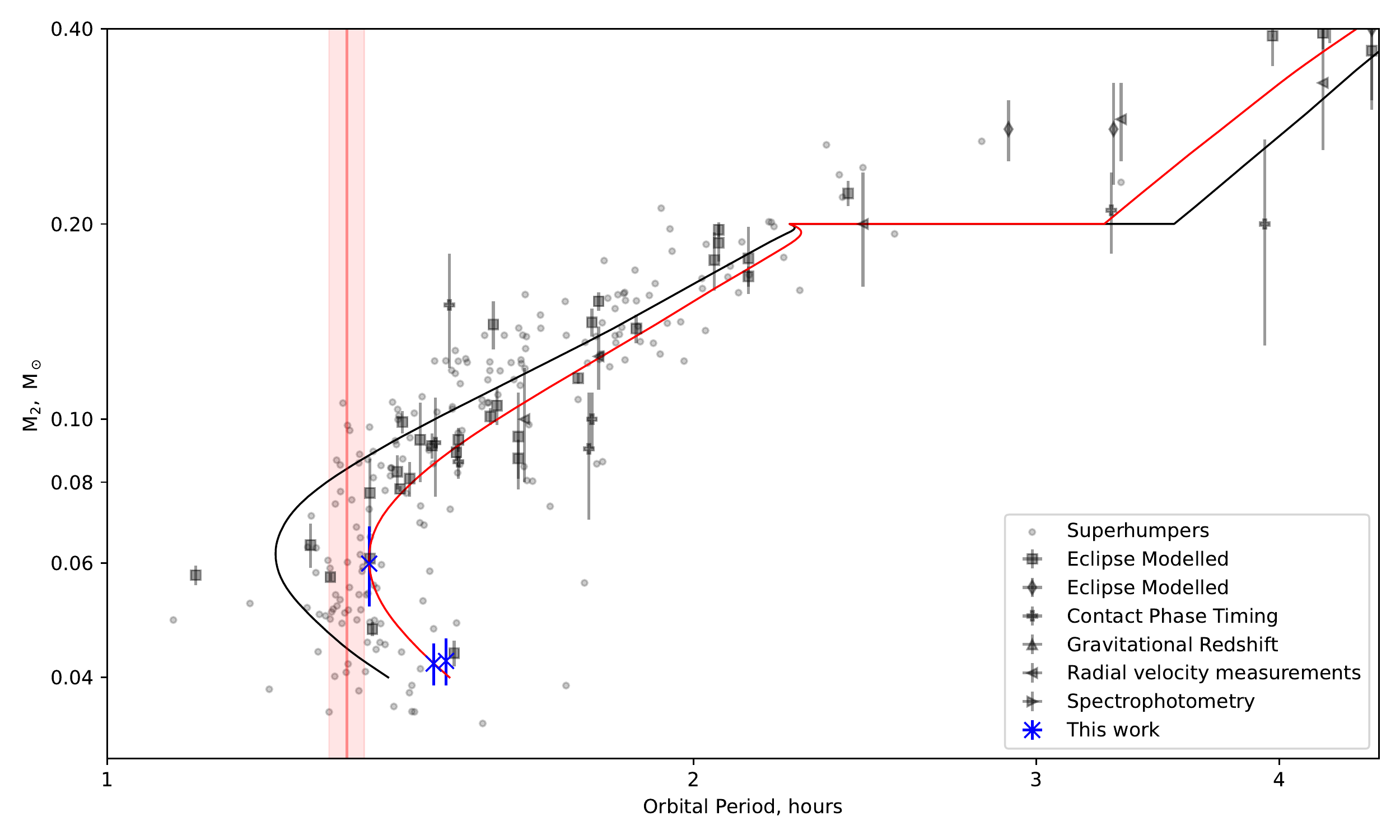}
    \caption{Donor evolution tracks -- note that both axes are scaled logarithmically. Solid black line: the standard donor sequence from \citet{knigge11}, solid red line: the `optimal' donor track from \citet{knigge11}. Vertical red line and shaded region: average period and $1\sigma$ region of these observed data between 76 and 82 minutes, $79.6\pm1.6$ minutes. Symbols denote the method used to analyse the data. \textbf{EM(U)}: Eclipse modelling with ULTRACAM, ULTRASPEC, or HiPERCAM data. \textbf{EM}: Eclipse modelling with other instruments. \textbf{CPT}: Contact phase timing. \textbf{GR}: Gravitational redshift. \textbf{RV}: Radial velocity measurement. \textbf{SM}: Spectrophotometric modelling. Blue crosses are the systems from this work.}
    \label{fig:M2_vs_P}
\end{figure*}

All three systems were candidate post-period minimum systems based on their periods and preliminary eclipse data; none show a prominent bright spot (indicative of a low mass transfer rate), or significant donor flux (implying a dim donor). 
As a result of this work, ASASSN-16kr and SSSJ0522-3505 are confirmed as having evolved through the period minimum and now have sub-stellar donors, and ASASSN-17jf lies in the period minimum region of Figure~\ref{fig:M2_vs_P}.
Additionally, all three white dwarf masses we derive in this analysis fall within the range of CV white dwarf masses observed by \citet{pala2020}, of $\langle M_{\rm WD}\rangle = 0.83 \pm 0.17 M_\odot$, significantly higher than the pre-CV DA white dwarf mass of only $0.66 \pm 0.15 {\rm M_\odot}$ \citep{mccleery2020}. Table~\ref{table:system_parameters} summarises the results for each system.

\subsection{Are we correct in assuming an unobscured white dwarf?}
\label{sect:impure white dwarf discussion}

As discussed in \S\ref{sect:white dwarf temperature report}, we believe the white dwarf colours could differ from model grids because the white dwarf ingress/egress is contaminated by an additional source of light, perhaps a boundary layer close to the surface.
If the eclipse we see is polluted by some other feature, our modelling will be wrong in two key elements: the colours we compare to model atmospheres will be incorrect, and the ingress and egress durations that constrain the white dwarf radius will not be accurate.
\citet{Spark2015} conducted a study into the validity of assuming a pure white dwarf, comparing CV eclipse observations with white dwarfs with and without a few types of surface features such as boundary layers on the white dwarf, hot spots, or an optically thick or thin equatorial belt.
These features are revealed by a departure from symmetry between the white dwarf ingress and egress, but care must be taken not to confuse the flickering component of the CV with the signature of surface features. 

Unfortunately, detecting a surface layer or hot spot on the white dwarf requires both a high time resolution and high signal-to-noise ratios. \citet{Spark2015} make use of SALTICAM data at a cadence of 0.15s, but our observations have a $\sim$3-4s exposure time and have lower signal-to-noise. We are unable to measure the eclipse precisely enough to make claims about the nature of the white dwarf's surface.
The three systems of this work are prime candidates to search for WD eclipse asymmetries, as the issue of flickering corrupting the white dwarf ingress/egress derivative is largely mitigated; all three have little to no flickering present. 
Future observations at higher cadence would open the possibility of examining the surfaces of these white dwarfs, though a large telescope will necessary due to the faintness of the systems.

\subsection{The hot white dwarf of SSSJ0522-3505}
\label{sect:SSSJ0522-3505 white dwarf temperature discussion}

The effective temperature of white dwarfs in short period CVs is typically $\sim10000$K \citep{pala2017a}, but our observed colours of SSSJ0522-3505 indicate a much hotter \teff\ of $\sim25000$K, which we believe to be accurate as the system's observations are dominated by the white dwarf flux, and show roughly the same eclipse depth in the $r', g'$, and $u'$ bands, that would not be consistent with a lower white dwarf temperature.

Our measured effective temperature could be wrong, either as a result of poor flux calibration (see \S\ref{sect:flux calibrating the lightcurve}) or because the ingress/egress fluxes do not represent the fluxes of the white dwarf photosphere, as discussed in section~\ref{sect:impure white dwarf discussion}. However, our measured temperature is  $\sim10000$\,K hotter than expected, and we do not believe these effects have introduced an error of this magnitude. As support for this, we note that \citet{pala2017a} find that white dwarf temperatures from UV spectroscopy typically agree with those measured from eclipse lightcurves to within $\sim1000$\,K. Therefore, we explore below reasons why the white dwarf temperature in SSSJ0522-3505 might be unusually hot, but note that UV spectroscopy to confirm the white dwarf temperature is highly desirable.

The white dwarf in a CV is thought to settle at an equilibrium temperature, where radiative heat loss is balanced with two energy sources: energy released by infalling material, and a low level of "simmering" nuclear fusion in the white dwarf envelope \citep{Townsley2003, Townsley2004}, but there are several reasons that this white dwarf may be temporarily out of equilibrium. 
There is no reason, though it is unlikely, that a CV cannot form from a main sequence star with a brown dwarf companion, to produce a young CV with a low-mass donor and a white dwarf  still cooling from its formation temperature.
Once the donor has reconnected with its Roche lobe, it would rejoin the normal CV evolution track and otherwise behave as a normal CV, with a normal accretion rate but a younger, hotter white dwarf than is typical.

A recent dwarf nova outburst was observed in this system in 2011, and could have produced a temporary boost to \teff. During these events, the disc enters a hot, optically thick state, and the infall rate onto the white dwarf is greatly increased \citep{osaki1996}, releasing a significant amount of energy and heating the white dwarf surface.
This is only the most recent \textit{observed} outburst, as there is a gap in observations between 2013 and 2019 during which any outburst events would have gone unrecorded. This may be important, as recent X-ray observations of another post period minimum system, OV Bootis \citep{Schwope2021}, shows that the WD temperature is increased to 23000K, 5 months after outburst, 9000K hotter than its \teff\ prior to outburst. The increase in temperature can be long lasting; detailed observations of GW Lib have shown its WD is still 3000K hotter than equilibrium 8 years post-outburst\citep{Szkody2016}.
Another possibility is a recent classical nova -- thermonuclear runaway in an accreted surface layer on the white dwarf -- which would temporarily heat the white dwarf beyond its equilibrium temperature \citep{starrfield2016}, giving the impression of a hotter white dwarf than expected, though a classical nova resulting in such a strong heating effect would be surprising.

If, however, we assume the white dwarf is in thermal equilibrium, \teff\ can be used to estimate the long-term accretion rate of the system \citep{townsley2009}.
If our modelled \teff\ of SSSJ0522-3505 is both accurate and driven by accretion, it would correspond to $\dot M_{\rm WD} = 6\pm2 \times 10^{-10} M_\odot {\rm yr^{-1}}$, compared to typical accretion rates of $\sim10^{-11} M_\odot {\rm yr^{-1}}$ for CVs in the post-period minimum regime \citep{pala2017a}. 
Whilst high, a mass accretion rate of $10^{-10} M_\odot {\rm yr^{-1}}$ is not incompatible with the presence of dwarf nova outbursts in SSSJ0522-3505, since a hot, optically thick accretion disc would require an accretion rate of order $10^{-8} M_\odot {\rm yr^{-1}}$ \citep{Hameury1998} to be stable on long timescales.

\subsection{Comments on the state of understanding AML in CVs}
\label{sect:discussion AML}

In order to qualitatively evaluate missing AML we examine the period \textit{excess}, $P_{\rm ex} = P_{\rm obs} - P_{\rm model}$, where $P_{\rm model}$ is the period predicted by the \citet{knigge11} track with only 1x gravitational braking below the period gap, interpolated across $M_{\rm donor}$. $P_{\rm obs}$ is the observed period.
To determine $P_{\rm ex}$ from an estimate of $P_{\rm obs}, M_{\rm donor}$, we sample from a Gaussian distribution based on the observed mean and standard deviation of $M_{\rm donor}$ and interpolate the evolutionary tracks to get a corresponding $P_{\rm model}$ distribution. As $P_{\rm model}$ is very sensitive to $M_{\rm donor}$, the $P_{\rm model}$ error dominates the uncertainty in $P_{\rm ex}$.
A positive $P_{\rm ex}$ tells us the model is missing AML, and a negative $P_{\rm ex}$ indicates a model that has too much AML, relative to an observation.

Our reported $P_{\rm ex}$ should be treated with caution, and is only provided as an illustrative parameter. The validity of $P_{\rm ex}$ is vulnerable on two key systematic biases; the validity of $P_{\rm model}$, and the inherent physical variation of the CV population. 

CVs may follow inherently different evolutionary tracks due to differences in donor metallicity \citep{stehle1997, harrison2016}, white dwarf mass \citep{knigge2006}, and the age of the donor when it first contacts the Roche lobe \citep{howell2001}. A population-wide scatter in this parameter space is not captured in the \citet{knigge11} model, which uses fixed values for these variables, but justification for the adopted values are given \citep{knigge11, knigge2006}.
If any individual system deviates from the adopted values in the models of \citet{knigge11} then $P_{\rm ex}$ for that system will be influenced by these differences as well as any extra AML. However, conclusions about $P_{\rm ex}$ drawn from the population at large should remain robust, as long as the population doesn't differ systematically from the values adopted in the models. The white dwarf mass used by \citet{knigge11} is somewhat lower than more recent observations suggest, using $M_{\rm WD} = 0.75 M_\odot$ versus the more recent value of $M_{\rm WD} = 0.83 \pm 0.17 M_\odot$ \citep{pala2020}. Modelling will be necessary to properly characterise the effect of this change on the donor evolutionary tracks, as this will affect both the size of the Roche lobes, and the rate of gravitational wave AML. However, other CV models suggest that the effect will be small; at most around 2 minutes \citep{goliasch2015}. 

More seriously, $P_{\rm ex}$ is only an accurate measure of additional AML, if the underlying donor physics in the model are correct. For example, if the models incorrectly predict the mass of systems in the period gap, this can have a large effect on $P_{\rm ex}$. In the models of \citet{knigge11} this mass is fixed at the empirically derived value of 0.2$M_\odot$. Observations of superhumping and eclipsing CVs suggest that period gap occurs at donor masses of $0.20 \pm 0.02 M_\odot$ \citep{knigge2006}. Using model tracks with lower or higher masses for the donor mass of the period gap would change the absolute value of $P_{\rm ex}$. However, the broad trend in $P_{\rm ex}$ will again be unchanged.

\begin{figure}
    \centering
    \includegraphics[width=\columnwidth]{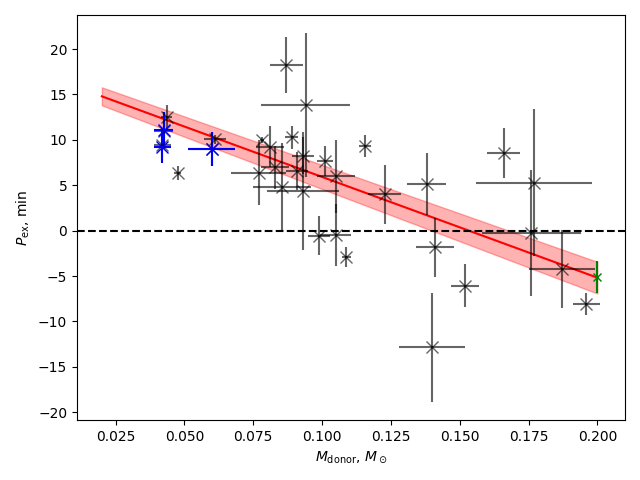}
    \caption{Showing period excess, $P_{\rm ex}$ (see \S\ref{sect:discussion AML}) against the \citet{knigge11} ``standard'' evolutionary track for short period CVs from Figure~\ref{fig:M2_vs_P}, excluding superhumpers, plotted as grey crosses. The three systems from this work are plotted as blue crosses. The solid red line shows the best-fit straight line to the plotted data, and the shaded red band shows the $1\sigma$ region of the fit. The green cross and error bar shows the predicted $P_{\rm ex}$ at $M_{\rm donor} = 0.2 M_\odot$. The horizontal black dashed line is a guide to show $P_{\rm ex} \equiv 0$.}
    \label{fig:period excess}
\end{figure}

The result is plotted in Figure~\ref{fig:period excess}. We fit the data with a straight line, and as the data have significant uncertainty in both axes, we minimise the sum orthogonal distance from the data \citep{hogg2010}. The best-fit parameters are a gradient of $-1.81+/-  0.13 M_\odot / {\rm hr}$, and a y-intercept of $0.283 \pm 0.016 {\rm\ hours}$. This gives $P_{\rm ex} = -4.89 \pm 1.81$ minutes at $M_{\rm donor} = 0.20M_\odot$, where a CV emerges from the period gap, roughly consistent with $P_{\rm ex}=0$, and the data show a clear increase in $P_{\rm ex}$\ towards lower $M_{\rm donor}$. 

We again stress that the only robust product of this analysis is the \textit{sign of the gradient} of the $M_{\rm donor} - P_{\rm ex}$ relationship, and that its steepness and y-intercept are both subject to systematic errors that we cannot capture in the statistical errors given above. Despite this, the clear and statistically highly significant increase in $P_{\rm ex}$ towards low masses implies that additional AML has a larger effect on the donor at low masses.

The strength of GWB falls with the total system mass, so we are left with three possibilities: the excess AML also declines in strength but more slowly than GWB; excess AML is roughly constant across the range of $M_{\rm donor}$; or excess AML actually increases in strength towards lower $M_{\rm donor}$. None of these options translate to the ``optimal'' \citet{knigge11} models which adopt additional AML of the same form as GWB.

We cannot convert our data to a more detailed AML prescription, as the donor radius and mass will be highly dependent on the mass loss \textit{history} of the system \citep{knigge11}. The donor star does not respond instantly to mass loss, but adjusts on a thermal timescale that is generally much longer than the mass loss timescale, so the degree of inflation a donor exhibits at a given mass will be affected by AML rates in the past. When a CV emerges from the period gap, the history is not significant as the donor has had ample time to adjust to the ``correct'' radius for its mass, but as it evolves to lower $M_{\rm donor}$, it will become more affected by the AML history of the system. 

It is not currently possible to distinguish between  proposed mechanisms for excess AML in CVs. However, an empirically determined, accurate AML prescription will help provide constraints for further exploration; the number of observations at the extremes of the donor track are now sufficient to begin to properly constrain the form of excess AML, but will require full evolutionary modelling with a focus on this aspect.

\section{Conclusion}

We contribute the component masses and radii, separations, white dwarf temperatures and surface gravities of three new short-period CVs to the population of well-characterised CV observations, two of which have extremely low-mass donor stars, and one which appears to be in the process of evolving through the period minimum. We measure the \teff\ of the white dwarf in SSSJ0522-3505 to be $\sim$10000K higher than is typical for a CV. We note that our derived temperature is quite uncertain, but we cannot confidently determine the origin of the discrepancy and summarise possible causes.
All three of the newly modelled systems lie within $1\sigma$ of the ``optimal'' model mass-radius evolutionary tracks from \citet{knigge11}.

The ``optimal'' tracks add an extra source of AML that takes the form of $1.5$ times the GWB. By examining the period excess between the growing set of observed CV donor radii and models, we demonstrate that this does not properly describe the missing AML. Rather than tracking the GWB as the CV evolves to lower masses, we find that the excess AML grows in strength relative to gravitational losses as the donor shrinks. The degree of inflation of the donor should provide an empirical diagnostic for this excess AML. Deriving a more quantitative AML prescription is beyond the scope of this work, as it would require fitting detailed evolutionary models to observations, due to the degree of donor inflation having a complex relationship with the AML history of the system.

\section*{Acknowledgements}

TRM acknowledges the support of the Science and Technology Facilities Council (STFC) grant STFC ST/T000406/1 and the Leverhulme Trust.

This work has made use of data from the European Space Agency (ESA) mission {\it Gaia} (\url{https://www.cosmos.esa.int/gaia}), processed by the {\it Gaia} Data Processing and Analysis Consortium (DPAC, \url{https://www.cosmos.esa.int/web/gaia/dpac/consortium}). Funding for the DPAC has been provided by national institutions, in particular the institutions participating in the {\it Gaia} Multilateral Agreement.

This research has made use of the NASA/IPAC Infrared Science Archive, which is funded by the National Aeronautics and Space Administration and operated by the California Institute of Technology.

We thank the anonymous referee for their careful reading of the paper, which greatly improved the quality of the results.

\section*{Data Availability}

The data underlying this article will be shared on reasonable request to the corresponding author.




\bibliographystyle{mnras}
\bibliography{period_bouncers_2020} 



\newpage
\clearpage
\appendix

\section{Photometric extraction and calibration}
\label{sect:photometric extraction and calibration}

The HiPERCAM data reduction pipeline \citep{dhillon2016} was used to perform debiassing and flat-field corrections on the raw frames. The software was also used for the extraction of aperture photometery, producing the flux in Analog-to-Digital Units, ADU, per frame of each source. 
A comparison star in the same image as the target was used to account for transparency variations, and standard stars from \citet{smith2002} were used to transform the lightcurves from ADU to the SDSS $u'g'r'i'z'$ photometric system.

\subsection{Calculating atmospheric extinction coefficients}
\label{sect:calcualting atmospheric extinction}

Atmospheric extinction was calculated using the longest continuous ULTRACAM observation available within 3 days of the target observations.
The atmospheric extinction values are reported in Table~\ref{table:atmos_extinction}.
No suitable observation was available in January 2020, so the average of the coefficients on 14 Oct 2018 and 30 Sep 2019 was used. 
Aperture photometery was extracted for five sources in these long observations, and the instrumental magnitude, $m_{\rm inst}$, vs airmass, $\chi$, was fit with a straight line for each source. 
The gradients of these lines are the atmospheric extinction coefficients, $k_{\rm ext}$, for the relevant band, and the y-intercept is the instrumental magnitude of that object above the atmosphere, $m_{\rm inst,0}$:
\begin{align*}
    m_{\rm inst} =& m_{\rm inst,0} + \chi k_{\rm ext} 
\end{align*}

\begin{table}
    \centering
    \caption{Atmospheric extinction coefficients for La Silla, derived from ULTRACAM/NTT observations.}
    \label{table:atmos_extinction}
    \begin{tabular}{cccc}
        \hline
        Date of Observation & Airmass Range & Band & $k_{ext}$ \\
        \hline
        \hline
        14 Oct 2018   & 1.30-1.98 & $u_{\rm reg}$ & $0.4476$ \\
                      &           & $g_{\rm reg}$ & $0.1776$ \\
                      &           & $r_{\rm reg}$ & $0.0861$ \\
        \hline
        30 Sept 2019  & 1.03-1.63 & $u_{\rm sup}$ & $0.4867$ \\
                      &           & $g_{\rm sup}$ & $0.1803$ \\
                      &           & $r_{\rm sup}$ & $0.0713$ \\
        \hline
    \end{tabular}
\end{table}

\subsection{Transformations between filter systems}
\label{sect:colour correction method}

The ULTRACAM photometric system previously matched the SDSS reasonably closely, however in early 2019 it was upgraded and now uses an SDSS-\emph{like} filter system with higher efficiency bandpasses, referred to as Super SDSS. There are three optical paths that are relevant:
\begin{itemize}
\item SDSS filters, $u', g', r', i', z'$;
\item ULTRACAM SDSS, NTT, $u_{\rm reg}, g_{\rm reg}, r_{\rm reg}, i_{\rm reg}, z_{\rm reg}$;
\item ULTRACAM Super SDSS,  NTT, $u_{\rm sup}, g_{\rm sup}, r_{\rm sup}, i_{\rm sup}, z_{\rm sup}$.
\end{itemize}

\begin{figure}
    \centering
    \includegraphics[width=\columnwidth]{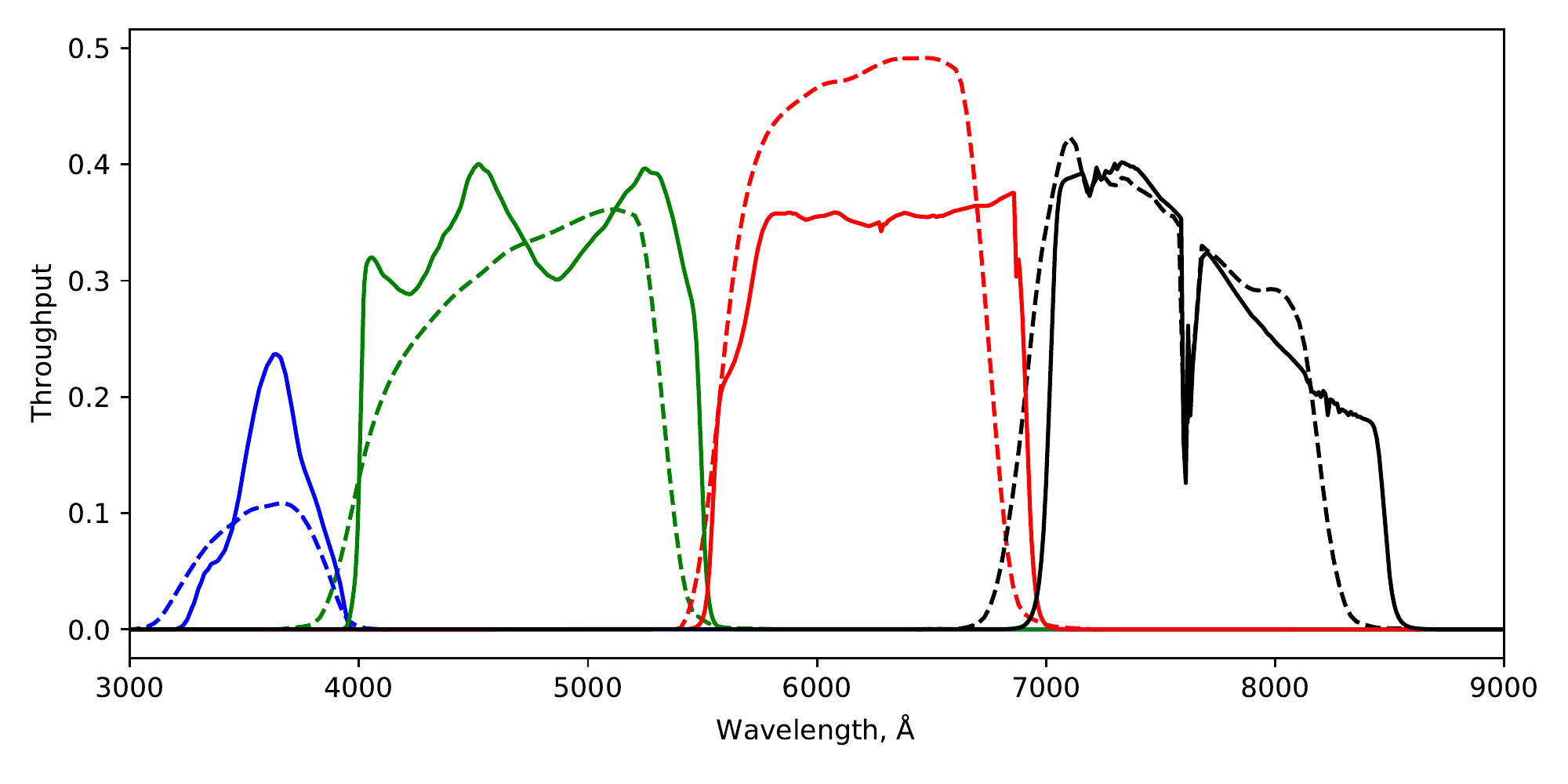}
    \caption{The differences in photometric throughput for SDSS filter system (dotted lines), and ULTRACAM Super SDSS filters, for ULTRACAM mounted on the NTT (solid lines). Blue: $u$ bands, Green: $g$ bands, Red: $r$ bands, Black: $i$ bands. Both throughputs include atmospheric extinction of $\chi = 1.3$.}
    \label{fig:sdss vs super filters}
\end{figure}

We aim to place our photometery in the SDSS $u'g'r'i'z'$ system, as this is the system later used by the white dwarf atmospheric models. The $u_{\rm reg}, g_{\rm reg}, r_{\rm reg}, i_{\rm reg}$\ filters were sufficiently similar to standard SDSS filters that the uncorrected magnitudes of standard reference stars from \citet{smith2002} could be used to calibrate absolute photometery without issue. However, with the new filters, there was concern that the different shape of the sensitivity curve, particularly in the $u'$\ band, differ enough from the standard filters to cause issues with our photometric calibration. Figure~\ref{fig:sdss vs super filters} illustrates the change in throughput between the SDSS photometric system, and the Super SDSS filters, on ULTRACAM on the NTT. 

To perform the colour corrections, the following equation for the magnitude of a star was used, using the $g'$\ band as an example:
\begin{equation}
    \label{eqn:gen magnitudes}
    g' = g_{\rm inst} + \chi k_{\rm ext} + g_{\rm zp} + c_{\rm g, sup}(g'-r') 
\end{equation}
where $g_{\rm zp}$\ is the zero point, $g_{\rm inst} = -2.5 \rm log(ADU/{\it t}_{\rm exp})$
for an exposure time of $t_{\rm exp}$, and $c_{\rm g, sup}$\ is the colour term correction gradient. 

The optical path of each system was simulated using the \texttt{pysynphot} package, with measured throughputs of all ULTRACAM components in the optical path. Models from \citet{Dotter2016} and \citet{Choi2016} were used to generate the \teff\ and \logg\ values of an $8.5$\ Gyr isochrone for main sequence stars with masses from 0.1 to 3 $M_\odot$. These span from \logg $= 3.73 \to 5.17$, and $\rm{T_{eff}} = 2900K \to 10,300K$. The Phoenix model atmospheres \citep{allard2012} were used to generate model spectra of each mass, which was then folded through each optical path to calculate an AB magnitude. In addition, white dwarf models with \logg$=8.5$\ were similarly processed \citep{koester2010, tremblay2009}, to asses the impact of the different spectral shape on the resulting colour terms.

\begin{figure*}
    \centering
    \includegraphics[width=0.9\textwidth]{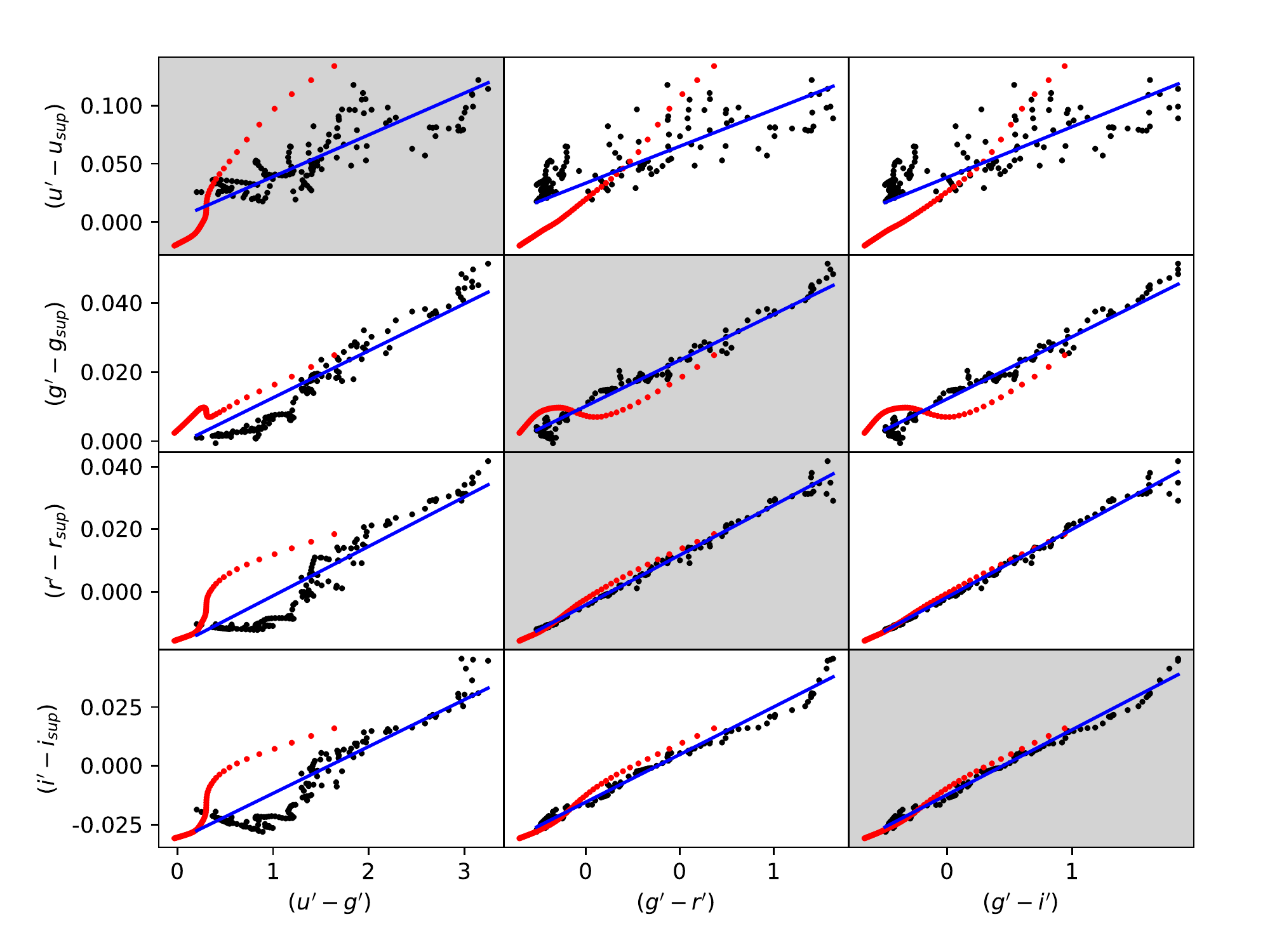}
    \caption{The difference between the classic SDSS photometric system, and the ULTRACAM SuperSDSS filters on the NTT, as a function of SDSS colours, are calculated for model atmospheres. Red points are Koester white dwarf models, black points are Phoenix main sequence model atmospheres, and the blue line is the best fit straight line to both datasets. When applying colour corrections, the highlighted relations were used.}
    \label{fig:all colour corrections}
\end{figure*}

We synthesised the colour terms between the SDSS and ULTRACAM Super SDSS systems, e.g., $g'-g_{\rm sup}$, for each model atmosphere. These data were plotted against SDSS colours, i.e. $(u'-g')$, $(g'-r')$, $(g'-i')$, and a straight line was fit to the colour relationship. In the example case of $g'-g_{\rm sup}$, this would be
\begin{align*}
    g' &= g_{\rm sup} + g_{\rm zp} + c_{\rm g, sup}(g'-r') \\
\end{align*}
Note we ignore the effects of secondary extinction. 
These relationships are shown in Figure~\ref{fig:all colour corrections} for all four ULTRACAM filters used to observe these CVs, and Table~\ref{table:all colour corrections} contains the coefficients of each colour term.
$(u'-g')$\ was used to correct $u$\ magnitudes, $(g'-r')$\ was used to correct $g$\ and $r$\ magnitudes, $(g'-i')$\ was used to correct the $i$\ band.
These colour corrections are not generally the same for main sequence stars and white dwarfs, though the colours of the white dwarfs presented in this work are all such that the discrepancy is on the order of a few percent, and is considered negligible.

\begin{table}
    \centering
    \caption{Colour term best fit lines from Figure~\ref{fig:all colour corrections}. The data are modelled by equations of the form $(u'-u_s) = \phi + c_u(u'-g')$, with $c_u$\ being the relevant colour gradient.}
    \label{table:all colour corrections}
    \begin{tabular}{cccc}
        Correction & Diagnostic &   y-intercept, $\phi$\   & Colour Gradient \\
        \hline
        \hline
          $(u'-u_s)$ &  $(u'-g')$   & 0.003 & 0.036 \\
                    &  $(g'-r')$   & 0.033 & 0.063 \\
                    &  $(g'-i')$   & 0.038 & 0.044 \\
        \hline
          $(g'-g_s)$ &  $(u'-g')$   & -0.001 & 0.014 \\
                    &  $(g'-r')$   & 0.010  & 0.027 \\
                    &  $(g'-i')$   & 0.012  & 0.018 \\
        \hline
          $(r'-r_s)$ &  $(u'-g')$   & -0.017 & 0.016 \\
                    &  $(g'-r')$   & -0.004 & 0.032 \\
                    &  $(g'-i')$   & -0.002 & 0.022 \\
        \hline
          $(i'-i_s)$ &  $(u'-g')$   & -0.031 & 0.020 \\
                    &  $(g'-r')$   & -0.015 & 0.040 \\
                    &  $(g'-i')$   & -0.012 & 0.028 \\
        \hline
        \hline
    \end{tabular}
\end{table}

\subsection{Calculating comparison star magnitudes}
\label{sect:comparison star mag calc}

Equation \ref{eqn:gen magnitudes} was used to calculate the zero points in each band from the standard star, for the SDSS photometric system.
The comparison star SDSS magnitudes are then determined. As the colour term corrections are dependent on SDSS colours, an iterative approach was used to converge on these values. The SDSS magnitudes are related to the instrumental magnitudes by:
\begin{align*}
    u' =& u_{\rm inst,0} + u_{\rm zp} + c_{\rm u, sup}(u' - g') \\
    g' =& g_{\rm inst,0} + g_{\rm zp} + c_{\rm g, sup}(g' - r') \\
    r' =& r_{\rm inst,0} + r_{\rm zp} + c_{\rm r, sup}(g' - r')
\end{align*}
Initially, $u',g',r'$\ magnitudes are set equal to the instrumental magnitudes, and a new set of $u',g',r'$\ magnitudes are calculated. The new values are then used to repeat the calculation until a new iteration produces no change, typically after $\sim$4 loops. For the data taken with $u_{\rm sup},g_{\rm sup},i_{\rm sup}$\ filters, the process is identical but replaces $r$ with $i$.

\subsection{Producing a flux-calibrated target lightcurve}
\label{sect:flux calibrating the lightcurve}

Finally, the target lightcurves can be calculated. We need to both correct the target star lightcurve for transparency variations, and convert from counts to calibrated fluxes. As we are producing a flux-calibrated lightcurve in the SDSS photometric system using a significantly different photometric system, the simple ADU ratio between the target and comparison is insufficient. Consider the target star $g'$ magnitude and flux, $g^t, F^t$, and comparison star $g'$ magnitude and flux, $g^c, F^c$:
\begin{align*}
    g^t =& g^t_{\rm inst,0} + g_{\rm zp} + c_{\rm g, sup}(g'-r')^t \\
    g^c =& g^c_{\rm inst,0} + g_{\rm zp} + c_{\rm g, sup}(g'-r')^c \\
\end{align*}
since,
\begin{equation*}
    g^t - g^c = -2.5{\rm log}\Big(\frac{F^t}{F^c}\Big) \\
\end{equation*}
we can write
\begin{align*}
    \frac{F^t}{F^c} =& 10^{-0.4(g^t_{\rm inst,0} - g^c_{\rm inst,0})} \cdot 10^{-0.4c_{\rm g, sup}\big((g'-r')^t - (g'-r')^c\big)} \\
    \frac{F^t}{F^c} =& \frac{ADU^t}{ADU^c}\cdot K^{t,c} \\
\end{align*}
where $K^{t,c} = 10^{-0.4c_{\rm g, sup}\big((g'-r')^t - (g'-r')^c\big)}$.
This accounts for differences in wavelength response between the two systems when calculating the flux ratio, and is applied to each frame. The $(g'-r')^t$\ magnitudes are calculated using a sigma-clipped mean instrumental magnitudes computed from all frames in the observation. In practice, the factor $K^{t,c}$\ varies from $\sim 1.0 - 1.1$\ across the three systems. 

ASASSN-16kr was observed in both the standard SDSS filters in 2018, and the super SDSS filters in 2019. This presented an opportunity to compare the corrected 2019 data with the fluxes observed in 2018. Additionally, both ASASSN-16kr and SSSJ0522-3505 use multiple standard stars across observations, which should agree if the calibration has been done correctly. In all cases, the flux-calibrated lightcurves were similar and the white dwarf colours consistent, suggesting an accurate flux calibration. See Appendix~\ref{appendix:lightcurves} for flux-calibrated lightcurves.

To account for residual error in flux calibration, we add a 3\% systematic error in quadrature to the white dwarf fluxes when fitting for the effective temperature.

\subsection{Ephemeris data}
\label{sect:ephemeris data}

ASASSN-16kr has existing ephemeris data in the literature \citep{kato2017}, whereas SSSJ0522-3505 and ASASSN-17jf were reported with tentative period estimates. These were used as starting points, and eclipse times from this work were used to refine the $T_\mathrm{0}$\ and $P$\ for all three systems. Only ULTRACAM eclipse timings were used to calculate the ephemerides in this paper.

To calculate the time of white dwarf mid-eclipse for each observation, the numerical derivative of the flux was fit with a a double-Gaussian model, as described in \citet{wood1985}.
Ideally, the derivative shows a negative peak at white dwarf ingress, and a symmetrical positive peak at egress, and each would be equidistant from the white dwarf mid-eclipse time, $T_\mathrm{ecl}$. By fitting the double-Gaussian model to a smoothed, numerical derivative of the lightcurve using a Markov Chain Monte Carlo (MCMC) method using a Gaussian process to evaluate the log-likelihood, we obtain $T_\mathrm{ecl}$ with uncertainties for each eclipse. These values are reported in Table~\ref{table:observations}.

For each observed $T_\mathrm{ecl}$, its eclipse number $N$\ (the number of eclipses since $T_0$) could unambiguously be determined from prior ephemeris data. 
An MCMC algorithm was used to fit a straight line model to the independent variable $N$\ and dependent variable $T_\mathrm{ecl}$, with a gradient $P$\ and intercept $T_0$. The model accounts for potential systematic differences in timing accuracy between instruments by also having variable error scale factors applied to all eclipses observed with a specific instrument, e.g. the timing reported for eclipses observed with ULTRACAM may be systematically offset from reality, and the errors associated with those observations might need to be larger than reported to be consistent with data from other instruments. The prior distribution assumed for these error factors was log-uniform ranging from 0.01 to 100, which favours the smallest factor consistent with the data. The values of $N$ for each system were chosen to minimise the covariance between $T_0$ and $P$.
The results of this ephemeris fitting are included in Table~\ref{table:system locations and ephemerides}.

\clearpage
\onecolumn
\section{Lightcurves}
\label{appendix:lightcurves}

\begin{figure}
    \centering
    \includegraphics[width=\columnwidth, trim={0 10cm 0 0}, clip]{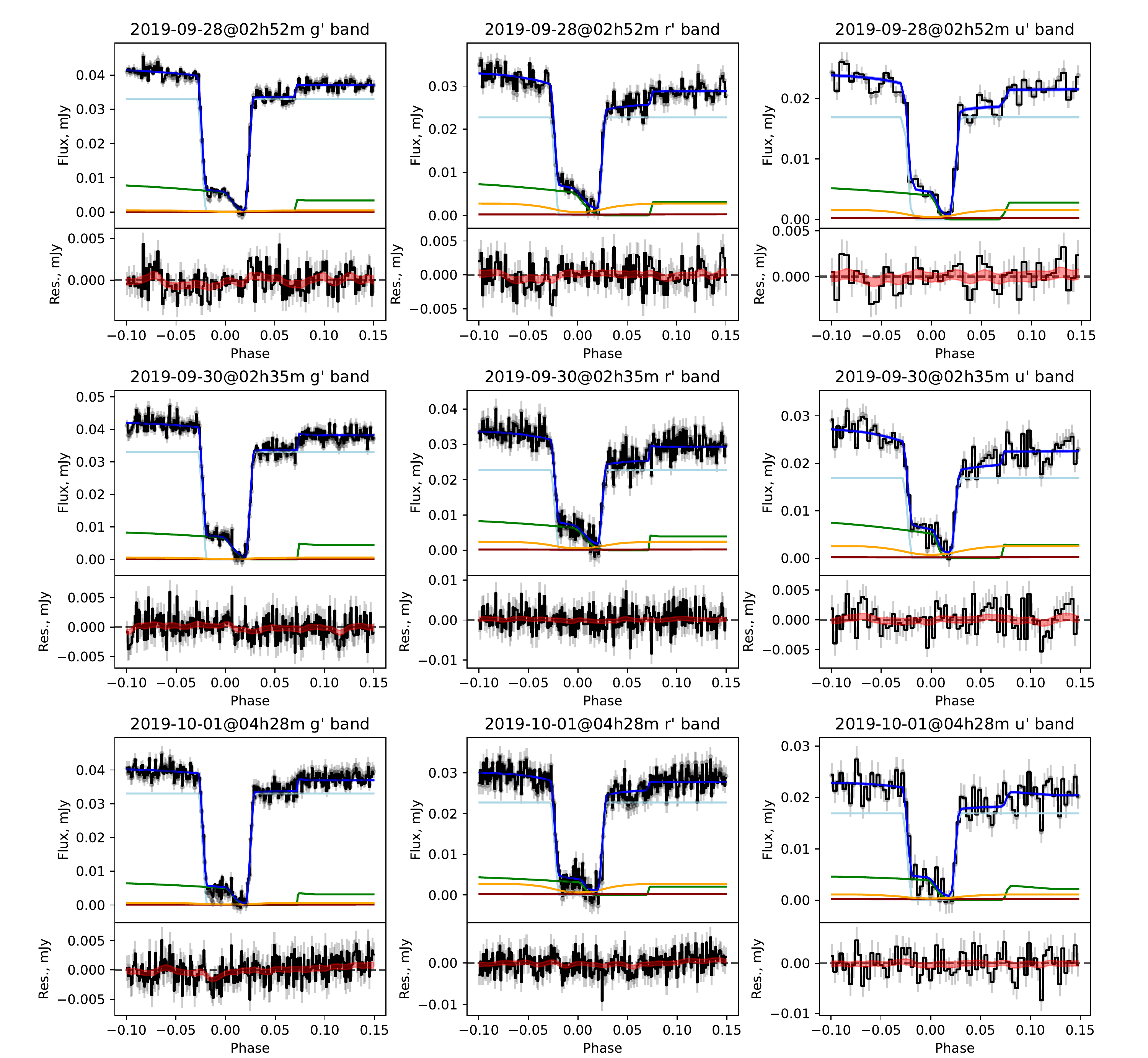}
    \caption{ASASSN-17jf lightcurve models. {\it Top}:~grey points are the observed flux; black line is the observed flux, with the mean Gaussian process sample subtracted; the dark blue line is the mean lightcurve model, and the blue band is the standard deviation on this in the MCMC chain. The components of the model are also shown: the light blue line is the white dwarf flux, green line is the bright spot, orange line is the disc, and the red line is the donor. {\it Bottom}:~The residuals between the data and model are plotted as the black line, with grey error bars. The Gaussian process 1-sigma region is shown as a red band.}
    \label{fig:ASASSN-17jf all lightcurves}
\end{figure}

\begin{figure}
    \centering
    \includegraphics[width=\columnwidth, trim={0 0cm 0 0}, clip]{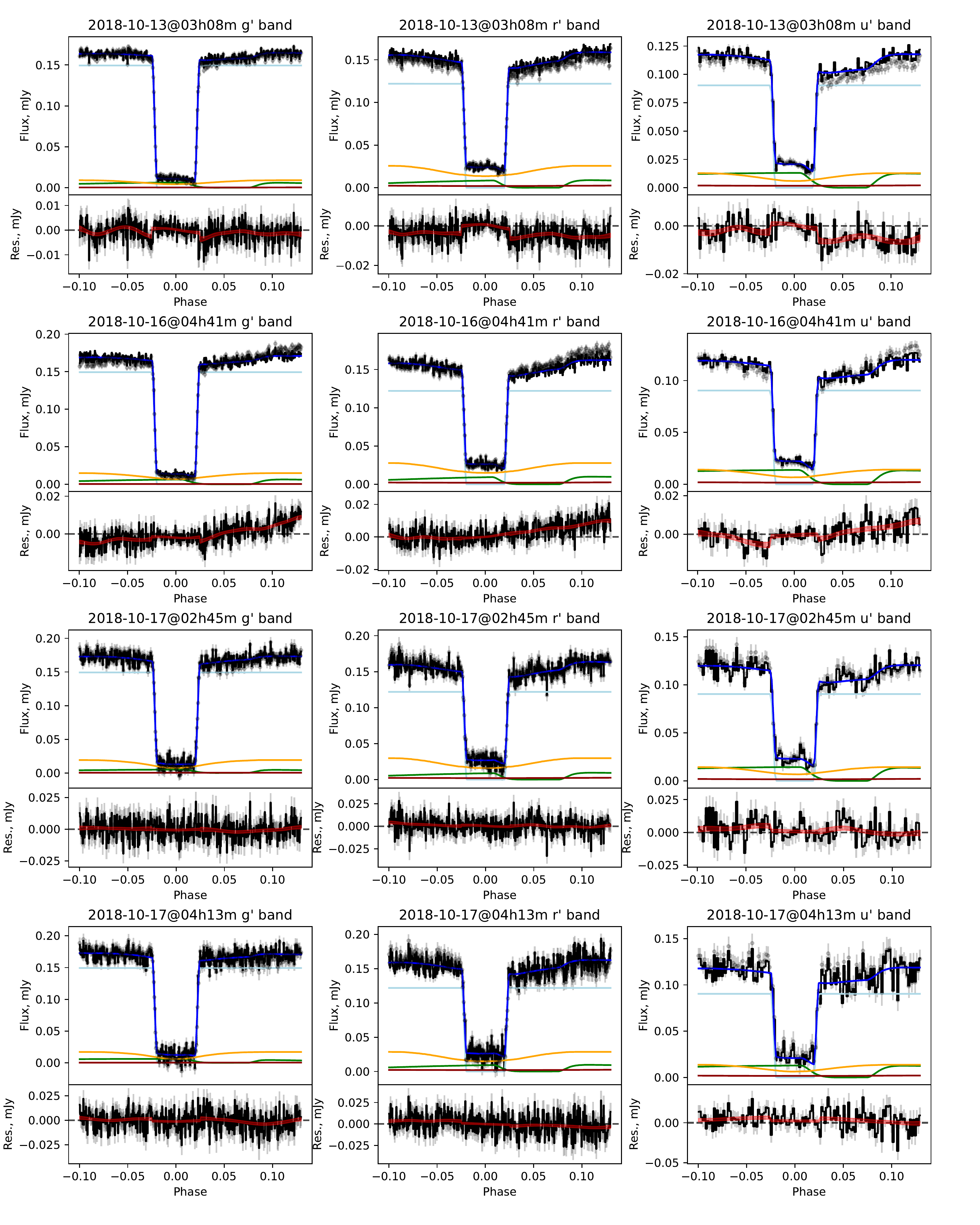}
    \caption{ASASSN-16kr lightcurve models. Symbols are the same as Figure \ref{fig:ASASSN-17jf all lightcurves}}
    \label{fig:ASASSN-16kr all lightcurves}
\end{figure}
\begin{figure}
    \centering
    \includegraphics[width=\columnwidth, trim={0 10cm 0 0}, clip]{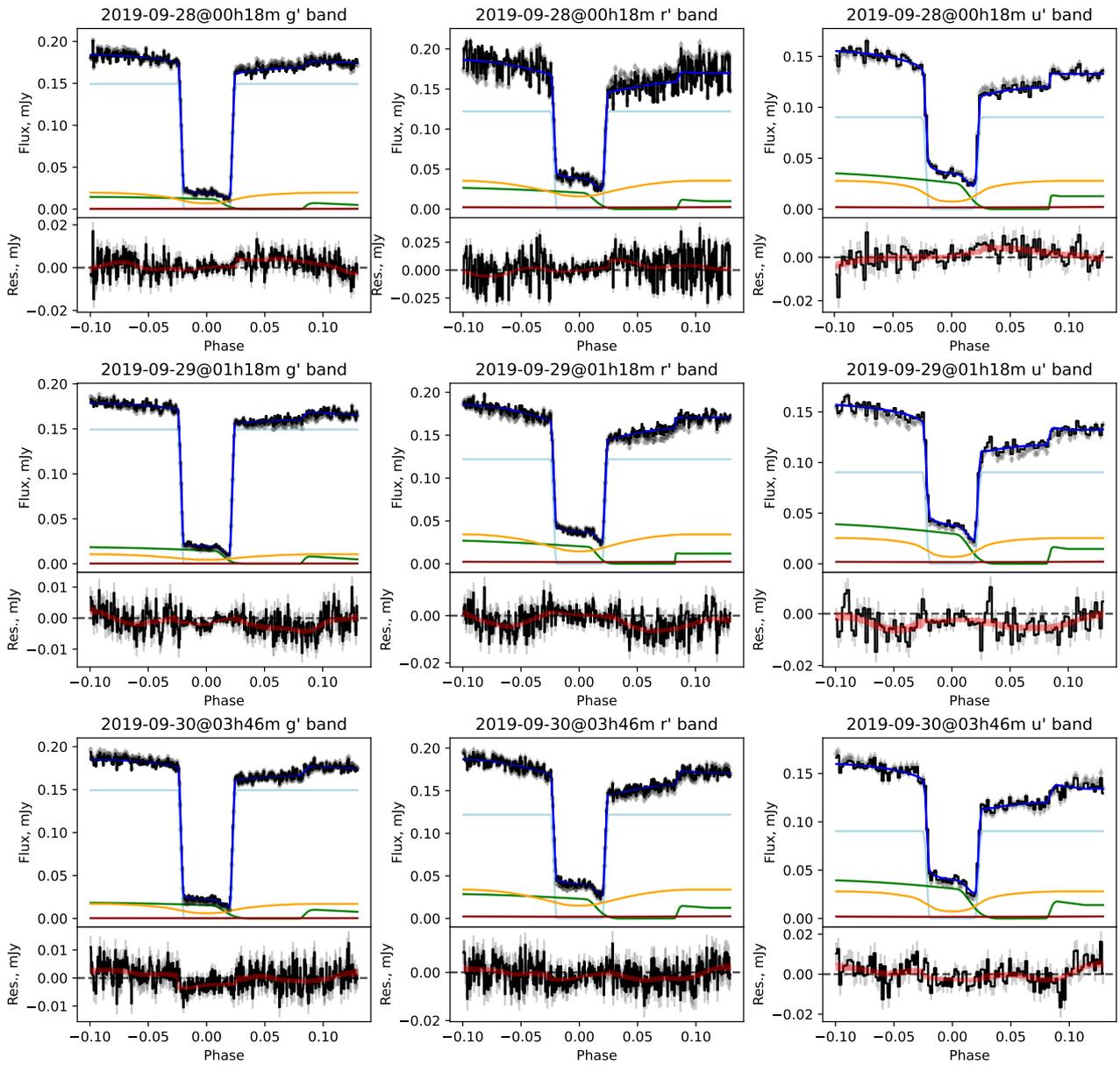}
    \caption{ASASSN-16kr lightcurve models (cont.)}
    \label{fig:ASASSN-16kr all lightcurves cont}
\end{figure}

\begin{figure}
    \centering
    \includegraphics[width=\columnwidth, trim={0cm 10cm 0cm 0cm}, clip]{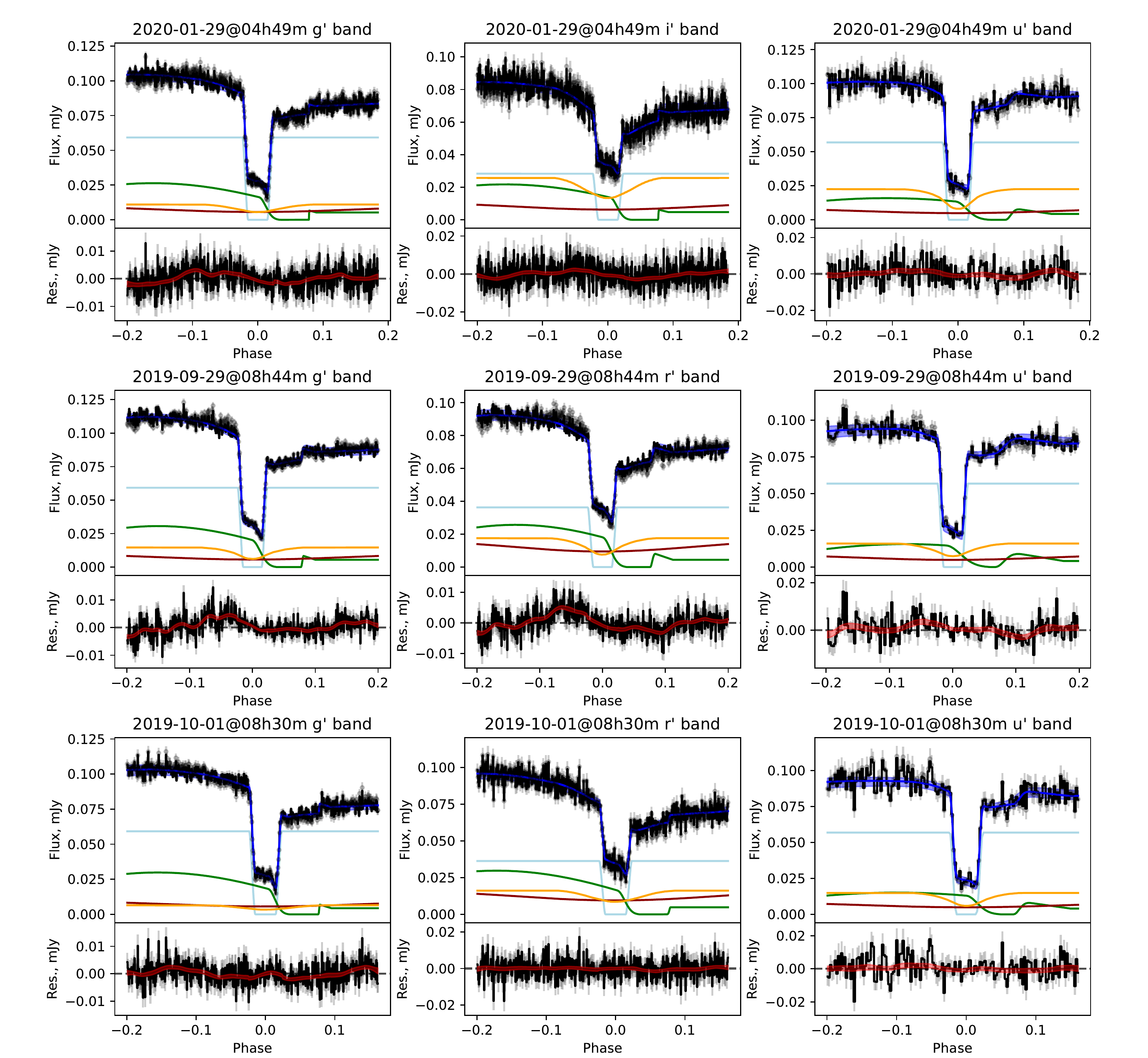}
    \caption{SSSJ0522-3505 lightcurve models. Symbols are the same as Figure \ref{fig:ASASSN-17jf all lightcurves}}
    \label{fig:SSSJ0522-3505 all lightcurves}
\end{figure}


\bsp	
\label{lastpage}
\end{document}